\pdfoutput=1

\documentclass[11pt,twoside,a4paper,cmspaper,final,collab]{cms-tdr}
\def\svnVersion{343181}\def\cmsCernNoTag{CERN-PH-EP/2015-213}\def\cmsCernDate{\today}\def\cmsMessage{Published in the Journal of High Energy Physics as \href{http://dx.doi.org/10.1007/JHEP11(2015)189}{\doi{10.1007/JHEP11(2015)189.}}}

\begin{document}\cmsNoteHeader{SUS-14-005}

\hyphenation{had-ron-i-za-tion}
\hyphenation{cal-or-i-me-ter}
\hyphenation{de-vices}

\RCS$Revision: 311383 $
\RCS$HeadURL: svn+ssh://svn.cern.ch/reps/tdr2/papers/SUS-14-005/trunk/SUS-14-005.tex $
\RCS$Id: SUS-14-005.tex 311383 2015-11-20 18:37:44Z smaruyam $

\providecommand{\mt}{\ensuremath{m_{\mathrm{T}}}\xspace}
\providecommand{\tauh}{\ensuremath{\PGt_\mathrm{h}}\xspace}
\newcommand{\mpt}{\ensuremath{\pt^{\text{miss}}}}
\providecommand{\NA}{\mbox{---}\xspace}

\providecommand{\PSGcmpDo}{\ensuremath{\widetilde{\chi}^\mp_{1}}\xspace}

\cmsNoteHeader{SUS-14-005}
\title{Search for supersymmetry in the vector-boson fusion topology in proton-proton collisions at $\sqrt{s} = 8$\TeV}

\date{\today}

\abstract{
The first search for supersymmetry in the vector-boson fusion topology is presented.
The search targets final states with at least two leptons, large missing transverse momentum, and two jets with a large separation in rapidity.
The data sample corresponds to an integrated luminosity of 19.7\fbinv of proton-proton collisions at $\sqrt{s} = 8$\TeV
collected with the CMS detector at the CERN LHC.
The observed dijet invariant mass spectrum is found to be consistent with the
expected standard model prediction. Upper limits are set on the cross sections for chargino and neutralino production with two associated jets,
assuming the supersymmetric partner of the $\tau$ lepton to be the lightest slepton and the lightest slepton to be lighter than the charginos.
For a so-called compressed-mass-spectrum scenario in which the mass difference between the lightest supersymmetric particle
$\PSGczDo$ and the next lightest, mass-degenerate, gaugino particles $\PSGczDt$ and $\PSGcpmDo$ is 50\GeV,
a mass lower limit of 170\GeV is set for these latter two particles.
}

\hypersetup{%
pdfauthor={CMS Collaboration},%
pdftitle={Search for supersymmetry in the vector-boson fusion topology in proton-proton collisions at sqrt(s) = 8 TeV},%
pdfsubject={CMS},%
pdfkeywords={CMS, physics, supersymmetry, SUSY}}

\maketitle

\section{Introduction}

With the successful operation of the CERN LHC, numerous results placing constraints on extensions to the standard model (SM) have been presented by the
ATLAS and CMS experiments.
In particular, in models of supersymmetry (SUSY)~\cite{PhysRevD.3.2415, JETP.13.1971, FERRARA1974413, WESS197439, PhysRevLett.49.970, BARBIERI1982343, PhysRevD.27.2359},
limits in excess of 1\TeV have been placed on the masses of the strongly produced gluinos and first- and second-generation
squarks~\cite{CMSGluinoLimits,ATLASGluinoLimits,Chatrchyan:2013fea,Khachatryan:2015lwa,Aad:2014pda, epjc/s10052-015-3518-2, JHEP09(2014)176, JHEP05(2015)078}.
In contrast, mass limits on the weakly produced charginos ($\widetilde{\chi}^{\pm}_{i}$) and neutralinos ($\widetilde{\chi}^{0}_{i}$), with much smaller
production cross sections, are much less severe. The limits for charginos and neutralinos are especially weak in so-called compressed-mass-spectrum
scenarios, in which the mass of the lightest supersymmetric particle (LSP) is only slightly less than the masses of other SUSY states.
The chargino-neutralino sector plays a crucial role in the connection between dark matter and SUSY: in SUSY models with R-parity
\cite{Farrar:1978xj} conservation, the lightest neutralino $\PSGczDo$ often takes the role of the LSP and is a dark matter candidate.

Previous LHC searches \cite{CMSEWK,Aad:2014nua} for electroweak chargino and neutralino production have focused on final states with one or more
leptons ($\ell$) and missing transverse momentum ($\ptvecmiss$), \eg, $\PSGcpmDo \PSGczDt$ pair production followed by
$\PSGcpmDo\to\ell \nu \PSGczDo$ and $\PSGczDt \to \ell \ell  \PSGczDo$, where $\PSGcpmDo$
($\PSGczDt$) is the lightest (next-to-lightest) chargino (neutralino), and where the LSP $\PSGczDo$ is presumed to escape
without detection leading to significant $\mpt$.
However, these searches exhibit limited sensitivity in cases where the $\PSGcpmDo$ and $\PSGczDt$ are nearly mass degenerate with the $\PSGczDo$.
The mass difference $\Delta m = m_{\PSGcpmDo} - m_{\PSGczDo}$ is a crucial parameter dictating the sensitivity of the analysis.
While the exclusion limits in Refs.~\cite{CMSEWK,Aad:2014nua} can be as large as $m_{\PSGcpmDo} < 720$\GeV for a massless $\PSGczDo$,
they weaken to only $\approx$ 100\GeV for $\Delta m < 50$\GeV.
The current searches also exhibit limited sensitivity to models with SUSY particles that decay predominantly to $\tau$ leptons, even for LSP masses near
zero, due to the larger backgrounds associated with $\tau$-lepton reconstruction compared to electrons or muons.

\begin{figure}[tbh!]
\begin{center}
  \includegraphics[width=0.45\textwidth]{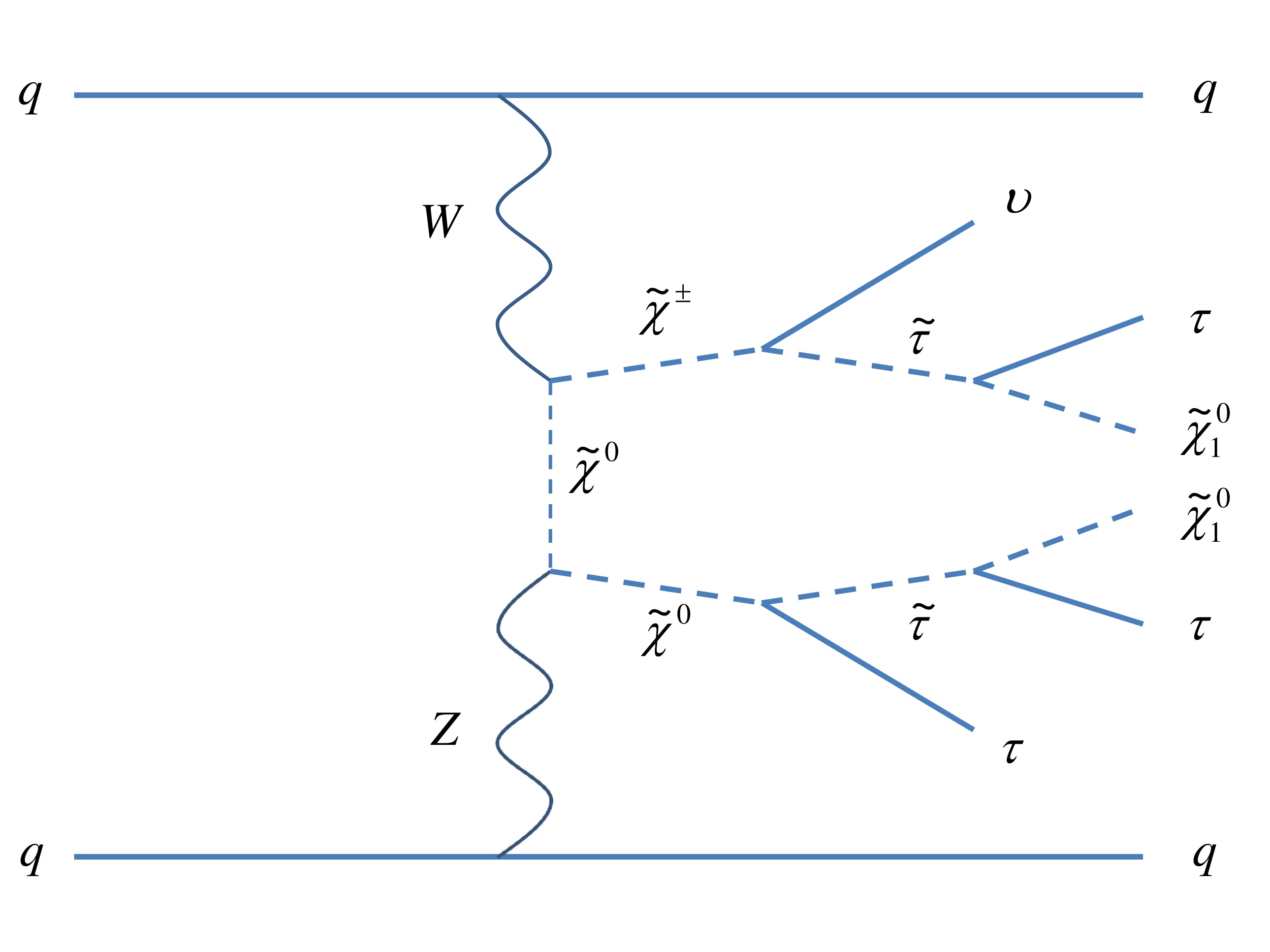}
  \includegraphics[width=0.45\textwidth]{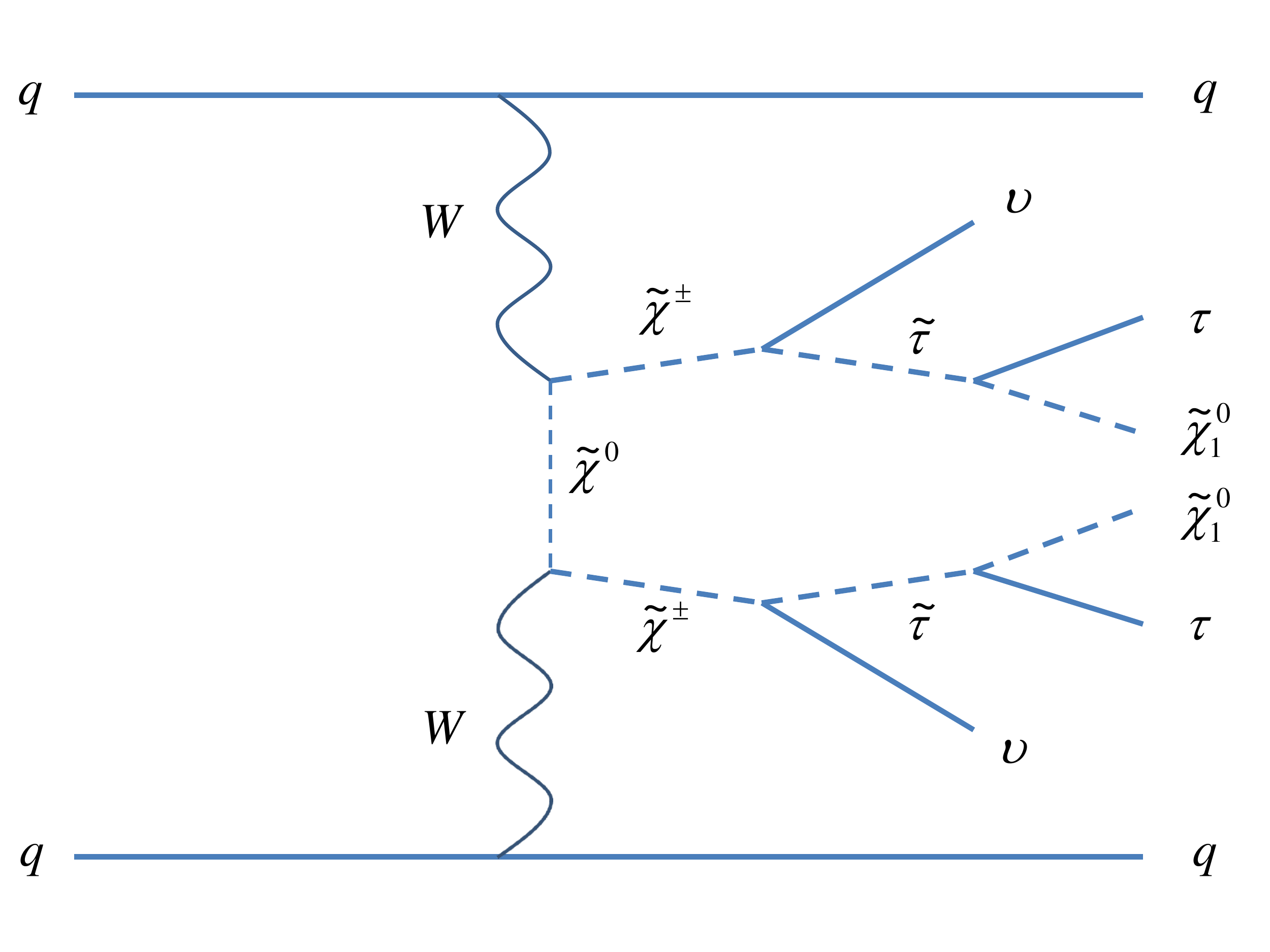}
\caption{Diagrams of (left) chargino-neutralino and (right) chargino-chargino pair production through vector-boson fusion followed by their decays to
leptons and the LSP $\PSGczDo$.}
\label{fig:feynVBF}
\end{center}
\end{figure}

Electroweak SUSY particles can be pair produced in association with two jets in pure electroweak processes in the vector-boson fusion (VBF)
topology \cite{VBFPheno}, which is characterized by the presence of two forward jets (\ie jets near the beam axis), in opposite hemispheres, leading to a large dijet invariant mass ($m_{jj}$).
Figure~\ref{fig:feynVBF} shows the Feynman diagrams for two of the possible VBF production
processes: chargino-neutralino and chargino-chargino production.

A search in the VBF topology offers a new and complementary means to directly probe the electroweak sector of SUSY, especially in compressed-mass-spectrum scenarios \cite{VBFPheno2}. It targets unexplored regions of SUSY parameter space, where other searches have limited sensitivity. It
differs fundamentally from the conventional direct electroweak SUSY searches mentioned above in that it uses the presence of jets with large transverse momenta (\pt) to suppress SM background. In this regard, it resembles searches for strongly produced SUSY particles.
However, unlike these latter studies, which present searches for the indirect production of charginos and neutralinos
through squark or gluino decay chains \cite{Chatrchyan:2013fea,Khachatryan:2015lwa,Aad:2014pda}, the VBF search does not require the
production of squarks or gluinos, whose masses might be too large to allow production at the LHC.

In this paper, we present a search for the electroweak production of SUSY particles in the VBF topology. The data, corresponding to an integrated
luminosity of 19.7\fbinv of proton-proton collisions at a centre-of-mass energy of $\sqrt{s}=8\TeV$, were collected with the CMS detector in 2012. Besides the two
oppositely directed forward jets ($j$) that define the VBF configuration, the search requires the presence of at least two leptons ($\Pe, \mu,$ or $\PGt$) and large \mpt. The events are classified into one of eight final states depending on the dilepton content and charges $\Pe\mu jj$, $\mu\mu jj$, $\mu\tauh jj$, and $\tauh\tauh jj$, where $\tauh$ denotes a hadronically decaying $\tau$ lepton and where we differentiate between final states with same-sign (SS) and opposite-sign (OS) dilepton pairs.
The dijet invariant mass distribution $m_{jj}$ is used to search for the SUSY signal.
Stringent requirements are placed on $\mpt$ and on the kinematic properties of the VBF dijet system to suppress SM background.
In particular, the R-parity conserving SUSY models we examine predict much higher average dijet \pt than is
typical for SM processes such as VBF Higgs boson production~\cite{CMS:2013bfa}, allowing us to suppress the background by a factor of $10^{2}$--$10^{4}$, depending on the background process.

The background is evaluated using data wherever possible.
The general strategy is to define control regions, each dominated by a different background process, through modification of the nominal selection
requirements. These control regions are used to measure the $m_{jj}$ shapes and probabilities for
background events to satisfy the VBF selection requirements. If the background contribution from a particular process
is expected to be small or if the above approach is not feasible, the $m_{jj}$ shapes are taken from simulation. In these cases, scale factors, defined as the ratio of efficiencies measured in data and simulation, are used to normalize the predicted rates to the data.

The paper is organized as follows. The CMS detector is described in Section \ref{sec:detector}. The
reconstruction of electrons, muons, $\tauh$ leptons, jets, and $\mpt$ is presented in Section \ref{sec:leptonRecoId}.
The dominant backgrounds and their simulated samples are discussed in Section \ref{sec:backgrounds},
followed by the description of the event selection in Section \ref{sec:event_selection}
and the background estimation in Section \ref{sec:bgestimation}.
Systematic uncertainties are summarized in Section \ref{sec:systematics}, and the results are presented in Section \ref{sec:results}.
Section 9 contains a summary.

\section{CMS detector}
\label{sec:detector}

The central feature of the CMS apparatus is a
superconducting solenoid of 6\unit{m} internal diameter, providing a magnetic field
of 3.8\unit{T}. Located within the solenoid volume are a silicon pixel and strip
tracker, a lead tungstate electromagnetic calorimeter (ECAL),
and a brass and scintillator hadron calorimeter. Muons are measured in
gas-ionisation detectors embedded in the steel flux-return yoke outside the solenoid.
Extensive forward calorimetry complements the coverage provided by the barrel and endcap detectors.
Forward hadron calorimeters on each side of the CMS interaction point cover the very forward angles of CMS,
in the pseudorapidity range $3.0 < \abs{\eta} <5.0$.

The inner tracker measures charged tracks with $\abs{\eta} < 2.5$
and provides an impact parameter resolution of $\approx$15\mum and a
transverse momentum resolution of about 1.5\% for 100\GeV charged particles.
Events are selected with a first-level trigger made of a system of
fast electronics, and a high-level trigger that consists
of a farm of commercial CPUs running a version of the offline
reconstruction optimized for fast processing. A detailed description of the CMS detector, along with a definition of the coordinate system and relevant kinematic variables, can be found in Ref.~\cite{CMS}.

\section{Object reconstruction and identification}\label{sec:leptonRecoId}

The missing transverse momentum vector $\ptvecmiss$ is defined as the projection on the plane perpendicular to the beam axis of the negative
vector sum of the momenta of all reconstructed particles in an event. Its magnitude is referred to as \mpt.
The jets and $\mpt$ are reconstructed with the particle-flow algorithm~\cite{pflow, CMS:2010byl}.
The anti-\kt clustering algorithm~\cite{antikt} with a distance parameter of 0.5 is used for jet clustering.
Jets are required to satisfy identification criteria designed to reject particles from
multiple proton-proton interactions (pileup) and anomalous behavior from the calorimeters.
For jets with $\pt > 30$\GeV and $\abs{\eta}< 2.5$ ($2.5 < \abs{\eta} < 5.0$),
the reconstruction-plus-identification efficiency is $\approx$99\% (95\%), while 90--95\% (60\%) of pileup jets are rejected~\cite{CMS-PAS-JME-13-005}.
The jet energy scale and resolution are calibrated through correction factors that depend on the \pt and $\eta$ of the
jet \cite{CMS:JetResol}. Jets originating from the hadronisation of bottom quarks (b quark jets) are identified using the loose working point of the combined
secondary vertex (CSV) algorithm~\cite{Chatrchyan:2012jua}, which exploits observables related to the long lifetime of b hadrons.
For jets with $\pt > 20$\GeV  and $\abs{\eta} < 2.4$, the probability of correctly
identifying a b quark jet is $\approx$85\%, while the probability of misidentifying a jet
originating from a light quark or gluon as a b quark jet is $\approx$10\%~\cite{CMS-PAS-BTV-13-001}.

Muons are reconstructed \cite{MUONreco} using the inner silicon tracker and muon detectors.
Quality requirements based on the minimum number of hits in the silicon
tracker, pixel detector, and muon detectors are applied to suppress
backgrounds from decays-in-flight and hadron shower remnants that reach the muon system.
Electrons are reconstructed \cite{Khachatryan:2015hwa} by combining tracks produced by the
Gaussian-sum filter algorithm with ECAL clusters.
Requirements on the track quality, the shape of the energy deposits in the ECAL,
and the compatibility of the measurements from the tracker and the ECAL
are imposed to distinguish prompt electrons from charged pions
and from electrons produced by photon conversions.
The electron and muon reconstruction efficiencies are $>$99\% for $\pt > 10$\GeV.

The electron and muon candidates are required to satisfy isolation criteria in order to reject
non-prompt leptons from the hadronisation process.
Isolation is defined as the scalar sum of the \pt values of reconstructed charged and neutral particles within a cone of radius $\Delta R =
\sqrt{\smash[b]{(\Delta\eta)^{2} + (\Delta\phi)^{2}}}=0.4$ around
the lepton-candidate track, divided by the \pt of the lepton candidate.
A correction is applied to the isolation variable to account for the effects of additional interactions. For charged particles, only tracks associated with the primary vertex are included in the isolation sums.
The primary vertex is the reconstructed vertex with the largest sum of charged-track $\pt^{2}$ values associated to it.
For neutral particles, a correction is applied by subtracting the energy deposited in the isolation cone by charged particles not
associated with the primary vertex, multiplied by a factor of 0.5. This factor corresponds approximately
to the ratio of neutral to charged hadron production in the hadronisation process
of pileup interactions.
In both cases, the contribution from the electron or muon candidate is removed from the sum and the value of the isolation variable is required to be
0.2 or less.

The muon identification-plus-isolation efficiency is 96\%
for muons with $\pt > 15$\GeV and $\abs{\eta} < 2.1$.
The rate at which pions undergoing $\pi^{\pm}\to\mu$ decay are misidentified as muons is $10^{-3}$ for pions
with $\pt > 10$\GeV and $\abs{\eta} < 2.1$.
The electron identification-plus-isolation efficiency is 85\% (80\%) for electrons with $\pt > 30$\GeV in the barrel (endcap) region~\cite{Khachatryan:2015hwa}.
The $j\to \Pe$ misidentification rate is $5\times 10^{-3}$ for jets with $\pt > 10$\GeV and $\abs{\eta} < 2.1$.

Hadronic decays of $\tau$ leptons are reconstructed and identified using the hadrons-plus-strips
algorithm \cite{TAU-11-001}, which is designed to optimize the performance of the $\tauh$ reconstruction by considering specific $\tauh$
decay modes.
To suppress backgrounds in which light-quark or gluon jets can mimic $\tauh$ decays, a $\tauh$ candidate is
required to be spatially isolated from other energy deposits in the event.
The isolation variable is calculated using a multivariate boosted-decision-tree technique
by forming rings of radius $\Delta R$ around the direction of the $\tauh$ candidate, using the energy deposits of particles not considered in the reconstruction of the $\tauh$ decay mode.
Additionally, $\tauh$ candidates are required to be distinguishable from electrons and muons in the
event by using dedicated criteria based on the consistency among the measurements in the tracker,
calorimeters, and muon detectors.
The identification and isolation efficiency is 55--65\% for a $\tauh$ lepton with $\pt > 20$\GeV and $\abs{\eta} < 2.1$,  depending on the \pt and $\eta$ values of the $\tauh$ candidate.
The rate at which jets are misidentified as a $\tauh$ lepton is 1--5\%, depending on the \pt and $\eta$ values of the $\tauh$ candidate.

The event selection criteria used in each search channel are summarized in Section~\ref{sec:event_selection} (see Table~\ref{tab:selectionSummary}).

\section{Signal and background samples}\label{sec:backgrounds}

The composition of SM background events depends on the final state and, in particular, the number of $\tauh$ candidates.
The most important sources of background arise from the production of $\PW$ or $\Z$ bosons in association with
jets ($\PW/\Z$+jets), and from $\ttbar$, diboson (VV: $\PW\PW$, $\PW\Z$, $\Z\Z$), and Quantum ChromoDynamics
(QCD) multijet events.
The \PW+jets events are characterized by
an isolated lepton from the decay of the W boson and uncorrelated jets misidentified as an \Pe, $\mu$, or $\tauh$.
Background from \PW+jets events is especially pertinent for final states with one $\tauh$ candidate.
Background from $\ttbar$ events usually contains one or two tagged b quark jets, in addition to a genuine isolated
\Pe, $\mu$, or $\tauh$.

Background from diboson events contains genuine, isolated leptons when the bosons decay leptonically, and jets that are misidentified as a $\tauh$ lepton when they decay hadronically.
The QCD background is characterized by jets that are misidentified as an \Pe, $\mu$, or $\tauh$ lepton.
The QCD multijet process is an appreciable background only for the $\tauh\tauh$ final states.

There are negligible contributions from background processes such as single-top and VBF production of a Higgs or Z boson.
These background yields are taken from simulation.

Simulated samples of signal and background events are generated using Monte Carlo (MC) event generators.
The signal event samples are generated with the
\MADGRAPH program (v5.1.5)~\cite{Alwall:2011uj}, considering pair production of $\PSGcpmDo$ and
$\PSGczDt$ gauginos ($\PSGcpmDo\PSGcpmDo$,
$\PSGcpmDo\PSGcmpDo$, $\PSGcpmDo\PSGczDt$, and
$\PSGczDt\PSGczDt$) with two associated partons.
The signal events are generated requiring a pseudorapidity gap $\abs{\Delta \eta} > 4.2$ between the two partons, with
$\pt > 30$\GeV for each parton.
Background event samples with a Higgs boson produced through VBF processes, and single top are generated with the \POWHEG program (v1.0r1380) \cite{Powheg}.
The \MADGRAPH generator (v5.1.3) is used to describe \Z{}+jets, \PW{}+jets, $\ttbar$, diboson, and VBF \Z boson production.

The MC background and signal yields are normalized to the integrated luminosity of the data.
The \ttbar background is normalized to the next-to-next-to-leading-logarithm level using the calculations of Refs.~\cite{Czakon:2013goa, PhysRevD.81.054028}.
The \Z{}+jets and \PW{}+jets processes are normalized to next-to-next-to-leading-order using the results from the \FEWZ v2.1~\cite{Gavin20112388} generator.
The diboson background processes are normalized to next-to-leading-order using the \MCFM v5.8~\cite{Campbell:2010ff} generator, while the VBF $\Z$ boson events are normalized to next-to-leading order using the {\sc{VBFNLO}} (v2.6)~\cite{2009CoPhC.180.1661A, 2011arXiv1107.4038B} program.
The single-top and VBF Higgs boson background yields are taken from the \POWHEG program, where the next-to-leading order effects are incorporated.
Signal cross sections are calculated at leading order using the \MADGRAPH generator.
All MC samples incorporate the  CTEQ6L1 \cite{CTEQL1} or CTEQ6M \cite{Nadolsky:2008zw} parton distribution functions (PDF).
The corresponding evaluation of uncertainties in the signal cross sections is discussed in Section~\ref{sec:systematics}.
The range of signal cross sections is 50--1\unit{fb} for $\PSGczDt=\PSGcpmDo$ masses of 100--400\GeV.
The \POWHEG and \MADGRAPH generators are interfaced with the \PYTHIA (v6.4.22) \cite{PYTHIA} program, which is used to describe the parton shower and hadronisation processes.
The decays of $\tau$ leptons are described using the {\TAUOLA} (27.1215)~\cite{TAUOLA_new} program.

The background samples are processed with a detailed simulation of the CMS apparatus using the {\GEANTfour} package \cite{Geant},
while the response for signal samples is modeled with the CMS fast simulation program \cite{FastSimulation}.
For the signal acceptance and $m_{jj}$ shapes based on the fast simulation, the differences with respect to the  \GEANTfour-based
results are found to be small ($<$5\%).
Corrections are applied to account for the differences.
For all MC samples, multiple proton-proton interactions are superimposed on the primary collision process, and events are
reweighted such that the distribution of reconstructed collision vertices matches that in data. The
distribution of the number of pileup interactions per event has a mean of 21 and a root-mean-square of 5.5.

\section{Event selection}
\label{sec:event_selection}
A single-muon trigger~\cite{MUONreco} with a \pt threshold of 24\GeV is used for
the $\mu\mu jj$, \Pe$\mu jj$, and $\mu \tauh jj$ final states. The $\tauh\tauh jj$ channels use a
double-$\tauh$ trigger~\cite{Zxsection} that requires \pt$>35$\GeV for each $\tauh$.
A requirement on pseudorapidity ($\abs{\eta} < 2.1$) is applied to select high quality and well-isolated leptons ($\Pe, \mu, \tauh$) within the tracker acceptance.
The \pt thresholds defining the search regions are chosen to achieve a trigger efficiency greater than 90\%.
For final states with at least one muon ($\mu \mu jj, \Pe\mu jj, \mu\tauh jj$),
events are selected by requiring a muon with $\pt > 30$\GeV.
For the $\tauh\tauh$ channels, both $\tauh$ candidates are required to satisfy $\pt>45$\GeV.

The following requirements are referred to as the ``central selection", and are applied to all final states.
Pairs of leptons are required to be separated by $\Delta R > 0.3$ and to originate from the primary vertex.
All channels require exactly two leptons satisfying selection criteria.
Events with an e or $\mu$ are required to have $\mpt > 75$\GeV,
while a requirement $\mpt > 30\GeV$ is used for the $\tauh\tauh jj$ final state
to compensate for the loss in acceptance due to the higher \pt threshold of $\tauh$ leptons while maintaining similar background rejection.
Background from $\ttbar$ events is reduced by removing events in which any jet has
\pt $>$ 20\GeV, is separated from the leptons by $\Delta R > 0.3$, and is identified as b-quark
jet using the loose working point of the CSV algorithm.

The ``VBF selection" refers to the requirement of at least two jets in opposite hemispheres ($\eta_{1} \eta_{2}<0$)
with large separation ($|\Delta\eta| > 4.2$).
Events are selected with at least two jets with $\pt > 50$\GeV and pseudorapidity $\abs{\eta} < 5.0$.
The $\mu^{\pm}\mu^{\pm}jj$ search region has a lower background rate with respect to other final states, which makes it
possible to relax the jet \pt requirement to 30\GeV to increase the signal acceptance.
The event selection criteria with $\pt > 30$\GeV are referred to as ``Loose``.
The event selection criteria with $\pt > 50$\GeV are referred to as ``Tight``.
Selected events are required to have a dijet candidate with $m_{jj} > 250$\GeV.

The signal region (SR) is defined as the events that satisfy the central and VBF selection criteria.
A summary of the event selection criteria used in each channel is presented in Table~\ref{tab:selectionSummary}.

\begin{table}[tbh!]
   \centering
\topcaption{Summary of the event selection criteria for the different final states.
The selections for the $\mu\mu jj$ and $\Pe\mu jj$ channels are presented in one column ($\ell_{\Pe/\mu} \mu jj$) as they are similar.
The symbol $\ell_{\Pe,\mu,\tauh}$ means that the lepton could be an electron, a muon, or a $\tauh$ lepton.
}

    \begin{tabular}{ c | l  c  c }
      \hline
      Selection &  \multicolumn{1}{c}{$\ell_{\Pe/\mu} \mu jj$}& $\mu\tauh jj$ & $\tauh\tauh jj$ \\
      \hline
      $\pt(\mu) [\GeVns{}]$                      & $\geq$30                  & $\geq$30  & \NA    \\
      $\pt(\ell_{\Pe/\mu}) [\GeVns{}]$           & ${\geq}15 (\Pe)$, ${\geq}10 (\mu)$ & \NA        & \NA     \\
      $\pt(\tauh) [\GeVns{}]$        & \NA                        & $\geq$20  & $\geq$45 \\
      $\abs{\eta(\ell_{\mu, \Pe, \tauh})}$       & $<$2.1                    & $<$2.1    & $<$2.1  \\
      $N^\text{b-tag}_{\text{jets}}$      & 0                          & 0          & 0        \\
      $\mpt [\GeVns{}]$                          & $>$75                     & $>$75     & $>$30   \\
      $\pt (\text{jets}) $                 & ${\geq}30/50$               & $\geq$50  & $\geq$30 \\
      $\abs{\eta(\text{jets})}$                & $\leq$5                    & $\leq$5    & $\leq$5  \\
      $\abs{\Delta\eta(\text{jets})}$          & $>$4.2            & $>$4.2  & $>$4.2  \\
      $\eta_{\text{1}} \eta_{\text{2}}$   & $<$0 & $<$0 & $<$0  \\
      \hline
    \end{tabular}
\label{tab:selectionSummary}
\end{table}

\section{Background estimation}
\label{sec:bgestimation}
The general methodology used to evaluate the background is the same for all final states.
We isolate various control regions (CR) to measure the VBF efficiencies (the probability for a given background component to satisfy the VBF selection criteria)
and $m_{jj}$ shapes from data, validate the modeling of the central selection criteria, and determine a correction factor to account for
the selection efficiency by assessing the level of agreement between data and simulation.
For each final state, the same trigger is used for the CRs as for the corresponding SR.
The VBF efficiency, measured in a CR satisfying only the central selection, is defined as the fraction of events in the CR
additionally passing the VBF event selection criteria.

The \ttbar, \PW+jets, and VV backgrounds are evaluated using the following equation:
\begin{equation}
N_{\mathrm{BG}}^{\text{pred}}(m_{jj}) = N_{\mathrm{BG}}^{\mathrm{MC}}(\text{central}) \, SF_{\mathrm{BG}}^{\mathrm{CR1}}(\text{central}) \,
\epsilon^{\mathrm{CR2}}_{\mathrm{VBF}}(m_{jj}),
\label{eq:bgsr}
\end{equation}
where $N_{\mathrm{BG}}^{\text{pred}}$ is the predicted background yield in the signal region,
$N_{\mathrm{BG}}^{\mathrm{MC}}\text{(central)}$ is the rate predicted by the ``BG" simulation (with BG = \ttbar, \PW+jets, or VV) for the central selection, $SF_{\mathrm{BG}}^{\mathrm{CR1}}(\text{central})$ is the data-to-simulation correction factor for the central region, given by the ratio of data to the ``BG" simulation in control region CR1, and $\epsilon_{\mathrm{VBF}}^{\mathrm{CR2}}$ is the VBF efficiency, determined as a function of $m_{jj}$ in data control sample CR2 or, in the case of VV events, from simulation.

The event selection criteria used to define the
CR must not bias the $m_{jj}$ distribution. This is verified,
in simulation and data, by comparing the $m_{jj}$ distributions with and without the selection criteria
used to define the CR. The background estimation technique used to measure the VBF efficiency and $m_{jj}$ shape
from data is performed with simulated events to test the closure, where closure refers to the ability of the method to predict the correct background yields when using simulation in place of data.
The closure tests demonstrate that the background determination techniques, described in detail below, reproduce the expected background distributions in both rate and shape to within the statistical uncertainties. The difference between the nominal MC background yields and the yields predicted from the closure test are added in quadrature with the statistical uncertainties of the prediction to define a systematic uncertainty.
Simulated samples are further used to verify that the composition of objects erroneously identified as leptons, and their kinematic properties, are similar between the CRs and SR, and thus that the scale factors $SF_{\mathrm{BG}}^{\mathrm{CR1}}(\text{central})$ are unbiased.
A variety of generators (\MADGRAPH, \PYTHIA, and \POWHEG) are used for this purpose to establish the robustness of this expectation.

\begin{figure}[tbh!]
  \centering
    \includegraphics[width=0.45\textwidth]{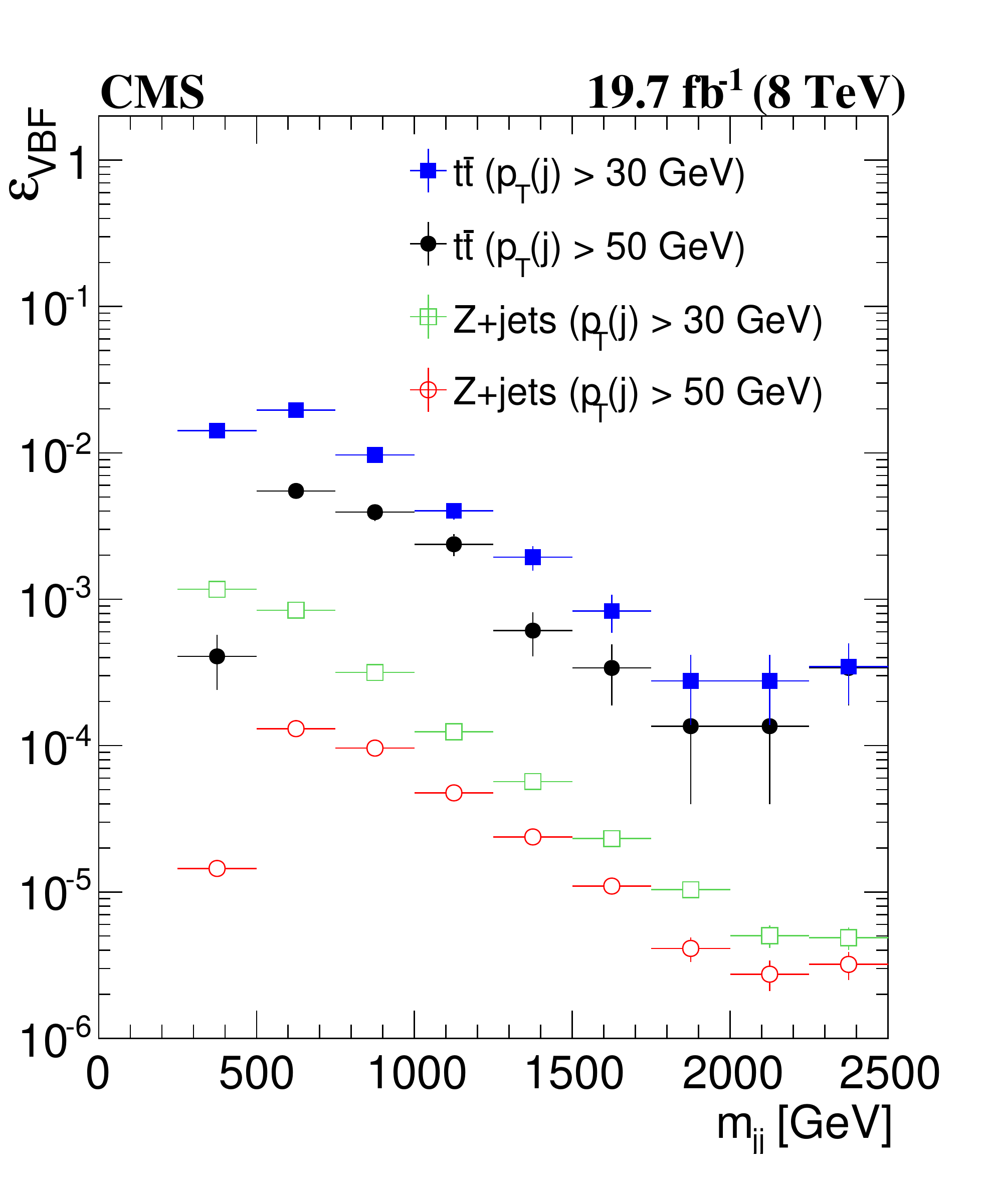}
  \caption{The VBF efficiency (see text) as a function of jet pair mass $m_{jj}$ measured for the \ttbar and \Z{}+jets control regions of the $\mu \mu jj$ final state, for the ``Loose" (\pt$>$ 30\GeV) and ``Tight" ($\pt>50$\GeV) event selections.
}
  \label{fig:muTauCRs_a}
\end{figure}

The production of $\ttbar$ events is an important source of background for the $\mu\mu jj$, $\Pe \mu jj$, and
$\mu\tauh jj$ final states. Control regions enriched with $\ttbar$ events are obtained by requiring the presence of at
least one reconstructed b-tagged jet with $\pt > 20$\GeV.
The \ttbar purity of the resulting data CR1s depends on the final state, ranging from 76 to 99\%.
The contributions of backgrounds other than \ttbar events are subtracted from the data CR1s using simulation.
The scale factors $SF_{\ttbar}^{\mathrm{CR1}}$ are then determined.
The uncertainties related to the subtraction procedure are propagated to the scale factors.
For the OS channels, the scale factors are consistent with unity to within 3\%.
The $\ttbar$ events with OS lepton pairs arise from genuine isolated leptons produced through leptonic $\PW$ boson decay, and are well modeled by the simulation.
On the other hand, \ttbar events with SS lepton pairs mostly contain a lepton candidate that is a misidentified hadron or jet, which is more difficult to accurately simulate. The scale factors for SS events range from 1.2 to 1.5, with statistical uncertainties up to 25\%. Since the fraction of lepton candidates that are in fact a misidentified hadron or jet varies according to the final state, the scale factors are determined independently for each channel.
In contrast, the VBF efficiency for a given combination of lepton flavors does not depend on the charge state, and thus each pair of final states with the same flavor combination shares the same VBF efficiency value.
The VBF efficiency is measured in data CR2 control samples obtained by additionally removing the charge requirement on the
lepton pair and relaxing or inverting the lepton isolation requirement (isolation sidebands) in order to enhance the purity and statistical precision of the CR2s.
Figure~\ref{fig:muTauCRs_a} shows the ``Tight" and ``Loose" VBF efficiencies measured from data, as a function of
$m_{jj}$, for events in the $\ttbar$ CR2s of the $\mu\mu jj$ final states.
The VBF efficiencies $\epsilon^{\mathrm{CR2}}_{\mathrm{VBF}}$ range from 0.02 to 0.003, with relative uncertainties below 11\% for $m_{jj} > 250$\GeV.
We verify that the b tagging, charge, and isolation requirements used to obtain the CR1 and CR2 samples do not bias the $m_{jj}$
shape or the kinematic distributions of the leptons.

The production of \PW+jets events presents an important source of background only for the $\mu\tauh jj$ search channels.
Samples enriched in \PW+jets events, with about 70\% purity according to simulation, are obtained by
requiring the transverse mass $\mt\equiv\sqrt{\smash[b]{2\mpt\pt^\mu[1-\cos(\Delta\phi_{\mu,\ptvecmiss})]}}$
 between \ptvecmiss and the muon transverse momentum $\pt^\mu$ to satisfy $40<\mt<110\GeV$.
The correction factor $SF^{\mathrm{CR1}}_{\PW\text{+jets}}$ is determined
to be $0.90 \pm 0.11$, where the uncertainty is a
combination of the statistical uncertainty from data, the statistical uncertainty from simulation, and the systematic uncertainty associated with the subtraction of the non-\PW+jets backgrounds from the data control sample.
The lepton and $\tauh$ isolation sidebands are used to obtain \PW+jets-enriched CR2 samples, with negligible expected signal contributions, not only to measure the VBF efficiencies and $m_{jj}$ shape from data, but also to provide further validation of the $SF^{CR1}_{\PW\text{+jets}}$ correction factor.
To validate the correction factor, the \PW+jets rate in the $\tauh$ isolation sideband is scaled
by $0.90 \pm 0.11$, and agreement between the data and the corrected  \PW+jets prediction from simulation is observed.
The VBF efficiency determined from the CR2 control sample is measured to be $\approx$1\% for $m_{jj}>250\GeV$.
Agreement between the VBF efficiencies of \Z{}+jets and \PW{}+jets processes is observed in the $\mu\tauh jj$ channel.

The background from VV events is significant for final states containing only electrons and muons,
comprising up to $10$\% of the total SM background in the OS channels, and up to $40$\% in the SS channels. The diboson background is
suppressed in the $\tauh$ final states because of the lower average \pt of the visible $\tau$-lepton decay products.
Diboson events have genuine isolated leptons and $\mpt$ from neutrinos and can satisfy the VBF selection when produced
in association with jets arising from initial-state radiation or from a SM VBF process.
We select diboson-enriched CR1 samples (97\% purity) by requiring at least three leptons and inverting the $\mpt$ requirement ($\mpt<75\GeV$).
The level of agreement between data and simulation for the event rates, VBF efficiencies, and $m_{jj}$ shapes are found to be the same for all types of VV processes in the CR1 samples.
The data-to-simulation correction factor is $SF^{\mathrm{CR1}}_{VV} = 1.12 \pm 0.06$.
The $m_{jj}$ distributions, following the VBF selections, are consistent between data and simulation.
Therefore, the VBF efficiency is taken directly from simulation.

The \Z{}+jets background is important for all OS final states.
This background is evaluated using the following relation:
\begin{equation}
  N_{\Z\text{+jets}}^{\text{pred}}(m_{jj}) = N_{\Z\text{+jets}}^{\mathrm{MC}}\text{(central)} \,
SF^{\mathrm{CR1}}_{\Z\text{+jets}}(\text{central}) \, SF^{\textrm{CR3}}_{\mpt} \, \epsilon^{\mathrm{CR1}}_{\mathrm{VBF}}(m_{jj}),
\label{equ:zjetsSR}
\end{equation}
where $N_{\Z\text{+jets}}^{MC}$, $SF_{\Z\text{+jets}}^{CR1}$, and $\epsilon_{VBF}^{CR1}$ have the same meaning as the analogous quantities in Eq.~(\ref{eq:bgsr}) (with BG = \Z{}+jets), and the $SF_{\mpt}^\mathrm{CR3}$ term is described below.
Control samples (CR1) dominated by $\Z \to \ell\ell$+jets events with $\ell=\Pe, \mu$ ($>98$\% purity) are selected by requiring $\mpt<75\GeV$ and an OS lepton pair mass $m_{\ell\ell}$ consistent with the $\Z$ boson ($60<m_{\ell\ell}<120\GeV$).
The rates and kinematic distributions of leptons in these control samples are consistent between the data and simulation.
These control samples are used to determine both the $SF_{\Z\text{+jets}}^{CR1}$ correction factors and the $\epsilon_{VBF}^{CR1}$ terms, in the same manner as described above for the analogous quantities.
The correction factors are unity to within 1\%.
Figure 2 shows the ``Tight`` and ``Loose`` VBF efficiencies measured from data, as a function of $m_{jj}$, for events in the $\mu\mu jj$ channel.
The VBF efficiencies range from $10^{-3}$ to $10^{-5}$.
The measured $\epsilon_{VBF}^{CR1}$ terms agree with the results from simulation within 23\%, which is taken as a systematic uncertainty both for the background prediction and for the VBF efficiency in simulated signal events.
Additional orthogonal \Z{}+jets control samples (CR3) are selected with similar selection criteria as used for the signal, maintaining the $\mpt$ requirement ($>$75\GeV) and inverting the VBF selection (\ie at least one of the VBF selection requirements is not satisfied: $\ge$2 jets, jet \pt, $\abs{\Delta\eta}$, or $\eta_{1} \eta_{2}$).
These control samples are used to determine the correction factors $SF^{CR3}_{\mpt}$, which account for differences between the data and simulation for high-$\mpt$ events.
The factors are $1.03 \pm0.03$ ($1.38 \pm 0.10$) for the $\tauh$ (light-lepton flavor) channels.
This mismodeling arises from the mismeasurement of $\pt$ for jets and leptons.

High-purity samples of $\Z\to \tau \tau \to \ell \tauh$ events, from which the $SF^{\mathrm{CR1}}_{\Z\text{+jets}}$ terms can be evaluated for the search channels with at least one $\tauh$, are obtained by removing the VBF selection and requiring
$\mt(\ell, \mpt)<15$\GeV.
The VBF efficiency for $\Z\to \tau \tau$+jets processes is obtained from data using the \Z$\to \ell \ell$+jets control samples described above.

The QCD multijet background is negligible for all final states except the $\tauh\tauh jj$ channel.
To estimate the QCD multijet contribution to this channel, we select a QCD-dominated ($>$90\% purity) CR1 by requiring two $\tauh$ candidates with relaxed isolation requirements.
In addition, we require that the CR1 events contain an SS $\tauh\tauh$ pair.
The SS signal region is thus included in CR1, but the potential impact of signal events is found to be negligible.
The QCD multijet background in the OS $\tauh\tauh$ channel is estimated by:
\begin{equation}
 N^{\text{pred}}_{\mathrm{QCD}}(m_{jj})=N_{\mathrm{QCD}}^{\mathrm{CR1}}\text{(central)} \, R_{\mathrm{OS/SS}} \,
\epsilon_{\mathrm{VBF}}^{\mathrm{CR2}}(m_{jj}),
 \label{eq:TauTauOS}
\end{equation}
where $N_{\mathrm{QCD}}^{\mathrm{CR1}}\text{(central)}$ is the yield observed in the CR1 sample with no VBF requirements, following subtraction of the non-QCD component from CR1 using simulation.
The OS-to-SS ratio $R_{\mathrm{OS/SS}}$ is obtained from a low-$\mpt$ ($\mpt<30$\GeV) region after subtraction of the non-QCD contributions: we find $R_{\mathrm{OS/SS}} = 1.33 \pm0.03$.
Besides its use in the background determination procedure [Eq.~(\ref{eq:TauTauOS})], the measured result for $R_{\mathrm{OS/SS}}$ is used to provide a cross-check: we use it to extrapolate from an SS to an OS control region, both selected requiring $\mpt<30$\GeV and two non-isolated $\tauh$ candidates.
The obtained prediction for the rate of non-isolated OS $\tauh$ leptons is in agreement with the observation.
Finally, the efficiency $\epsilon_{\mathrm{VBF}}^{\mathrm{CR2}}$ is measured in exclusive sidebands fulfilling inverted $\tauh$ isolation criteria.
It is estimated as the rate of events with two non-isolated $\tauh$ candidates plus two
jets satisfying the VBF requirements divided by the rate of events with two non-isolated $\tauh$ candidates
without any additional jet requirements. The measured efficiency is
$\epsilon_{\mathrm{VBF}}=0.35\%\pm 0.08\%\stat\pm0.06\%\syst$.

The QCD multijet background in the SS $\tauh\tauh jj$ channel is estimated using the following relation:
\begin{equation}
 N^{\text{pred}}_{\mathrm{QCD}}=N^{\text{fail-VBF}}_{\mathrm{QCD}}\, \frac{\epsilon_{\mathrm{VBF}}^{\text{non-iso}
\tauh}}{1-\epsilon_{\mathrm{VBF}}^{\text{non-iso} \tauh}}.
\label{eq:TauTauSS}
\end{equation}
Here, $N^{\text{fail-VBF}}_{\mathrm{QCD}}$ is the observed yield in data, with non-QCD background from simulation subtracted, in an SS $\tauh\tauh$ control sample requiring at least two jets not associated with one of the $\tauh$ candidates to fail any of the $\abs{\Delta \eta}$, $\eta_{1} \eta_{2}$, or $m_{jj}$ requirements.
The VBF efficiency $\epsilon_{\mathrm{VBF}}^{\text{non-iso} \tauh}$ ($\epsilon_{\mathrm{VBF}}$ for short) is measured in six exclusive $\tauh$ isolation sidebands,
without a $\mpt$ requirement and at least two jets.
The validity of the method is demonstrated in data by the agreement that is observed, within statistical uncertainties, of these six independent measurements of $\epsilon_{\mathrm{VBF}}$.
The six corresponding control samples in simulation are used to test the stability of $\epsilon_{\mathrm{VBF}}$ as a function of $\mpt$ and the $\tauh$ isolation requirements.
For this purpose, the probability for a single jet to be
misidentified as a $\tauh$ lepton is determined from simulation.
The misidentification rates depend on the jet \pt and are used to determine an overall event weight by randomly selecting two jets in the event to represent the $\tauh$ leptons.
The VBF efficiencies in simulation are calculated from these weighted samples and demonstrate consistency with respect to the different $\mpt$ and $\tauh$
isolation requirements at the level of $\approx$19\%, which is assigned as a systematic uncertainty in the background prediction.
The VBF efficiency for $m_{jj} > 250\GeV$ is $\epsilon_{\mathrm{VBF}}=6.7\%\pm0.5\%\stat^{+1.2\%}_{-0.5\%}\syst$.

\section{Systematic uncertainties}
\label{sec:systematics}

The main contributions to the total systematic uncertainty in the background predictions arise from the closure tests and
from the statistical uncertainties associated with the data control regions used to determine the
$\epsilon^{\mathrm{CR}}_{\mathrm{VBF}}$,
$SF^{\mathrm{CR1}}_{\mathrm{BG}}\text{(central)}$,
and $R_{\mathrm{OS/SS}}$ factors.
The relative systematic uncertainties in $SF^{\mathrm{CR1}}_{\mathrm{BG}}$ and $R_{\mathrm{OS/SS}}$ related to the statistical precision in the CRs range between 1 and 25\%, depending on the background component and search channel. For $m_{jj} > 250$\GeV, the statistical uncertainties in
$\epsilon^{\mathrm{CR}}_{\mathrm{VBF}}$ lie between 3 and 21\%, while the systematic uncertainties evaluated from the closure tests and cross-checks with data range from 2 to 20\%.
For the background $\epsilon^{\mathrm{CR}}_{\mathrm{VBF}}$, we assign no uncertainty due to the jet energy correction, as the $m_{jj}$ distributions are taken directly from the data control regions.

Less significant contributions to the systematic uncertainties arise from contamination by non-targeted background sources to the CRs used to measure $\epsilon^{\mathrm{CR}}_{\mathrm{VBF}}$, and from the
uncertainties in $SF^{\mathrm{CR1}}_{\mathrm{BG}}\text{(central)}$ due to the lepton identification
efficiency, lepton energy and momentum scales, $\mpt$ scale, and trigger efficiency.

The efficiencies for the electron and muon trigger, reconstruction, identification, and isolation requirements are measured with the ``tag-and-probe"
method~\cite{MUONreco,Khachatryan:2015hwa} with a resulting uncertainty of 2\%. The
$\tauh$ trigger and identification-plus-isolation efficiencies are measured
from a fit to the $Z\to\tau\tau\to\mu\tauh$ visible mass distribution in a sample selected with a single-muon trigger,
leading to a relative uncertainty of 4\% and 6\% per $\tauh$ candidate, respectively~\cite{Zxsection}.
The $\mpt$ scale uncertainties contribute via the jet energy scale (2--10\% depending on $\eta$ and \pt) and unclustered energy scale
(10\%) uncertainties, where ``unclustered energy" refers to energy from a reconstructed object that is not assigned to a jet with $\pt >10\GeV$ or to a lepton with $\pt >10\GeV$.

Since the estimate of the background is partly based on simulation, the signal and background rates are affected by similar sources of systematic uncertainty, such as the luminosity uncertainty of 2.6\%~\cite{CMS-PAS-LUM-13-001}.
The uncertainties in the lepton identification efficiency, lepton energy and momentum scale, $\mpt$ scale, and
trigger efficiency also contribute to the systematic uncertainty in the signal.

The signal event acceptance for the VBF selection depends on the reconstruction and identification
efficiency and jet energy scale of forward jets.
The jet reconstruction-plus-identification efficiency is $>$98\% for the
entire $\eta$ and $\pt$ range, as is validated through the agreement observed between data and simulation in the
$\eta$ distribution of jets, in particular at high $\eta$, in control samples enriched with $\ttbar$
background events. The dominant uncertainty in the signal acceptance is due to the modelling of the kinematic properties of jets, and thus the VBF efficiency, for forward jets in the
\MADGRAPH simulation. This is investigated by comparing the predicted and measured $m_{jj}$ spectra in the \Z{}+jets CRs.
The level of agreement between the predicted and observed $m_{jj}$ spectra is better than 23\%, which is assigned
as a systematic uncertainty in the VBF efficiency for signal samples.
The uncertainty in the signal acceptance due to the PDF set included in the simulated samples is evaluated in accordance with the PDF4LHC recommendations \cite{Alekhin:2011sk, Botje:2011sn} by comparing the results obtained using the
CTEQ6.6L, MSTW08, and NNPDF10  PDF sets~\cite{Nadolsky:2008zw,mrst2006,nnpdf10} with those from the default PDF set (CTEQ6L1).
The dominant uncertainties that contribute to the $m_{jj}$ shape variations include the \mpt and jet energy
scale uncertainties.
Correlations of the uncertainty sources are discussed in Section \ref{sec:results}.

\section{Results and interpretation}
\label{sec:results}
\begin{figure}[tbh!]
  \centering
    \includegraphics[width=0.45\textwidth]{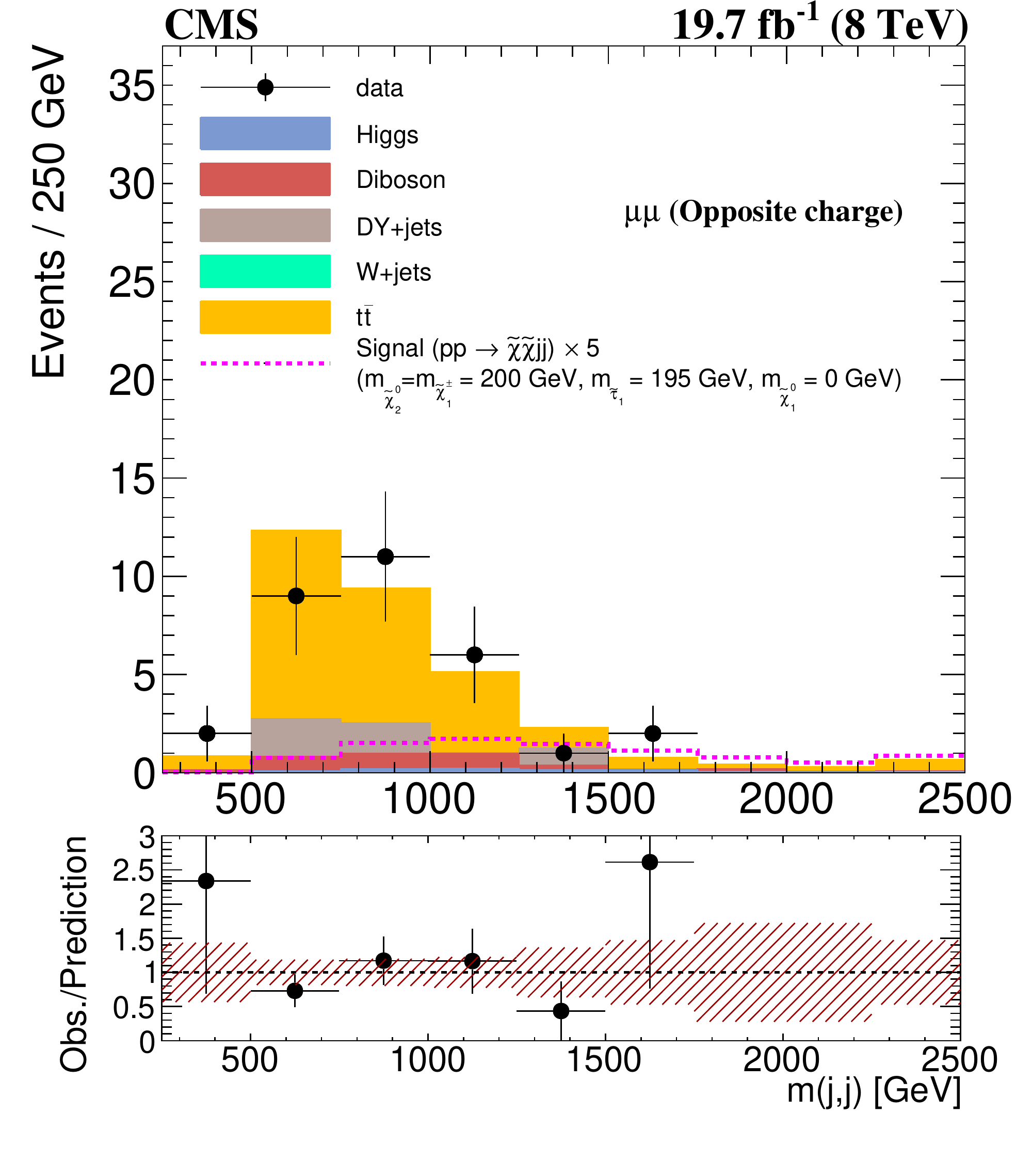}
    \includegraphics[width=0.45\textwidth]{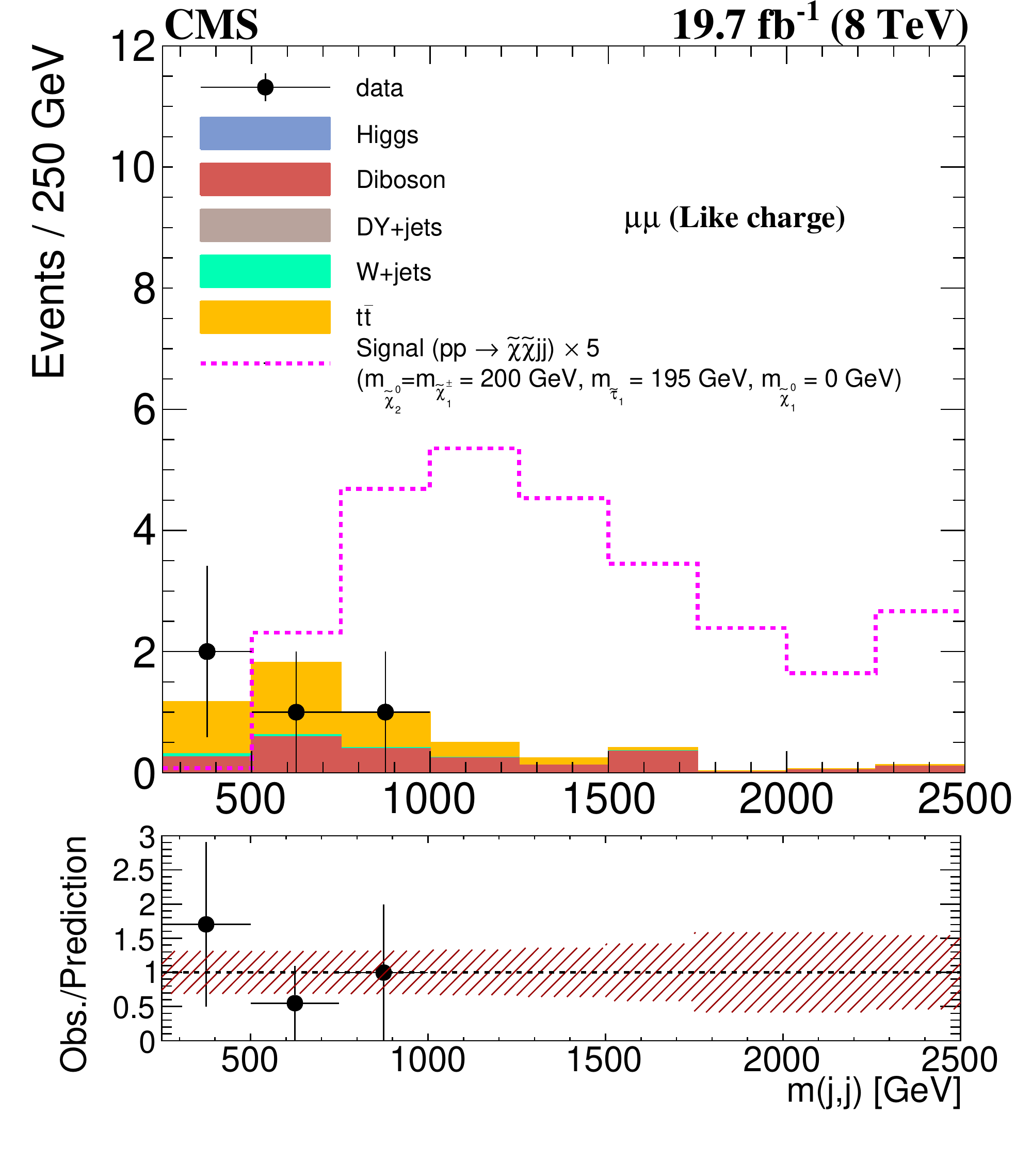}
    \includegraphics[width=0.45\textwidth]{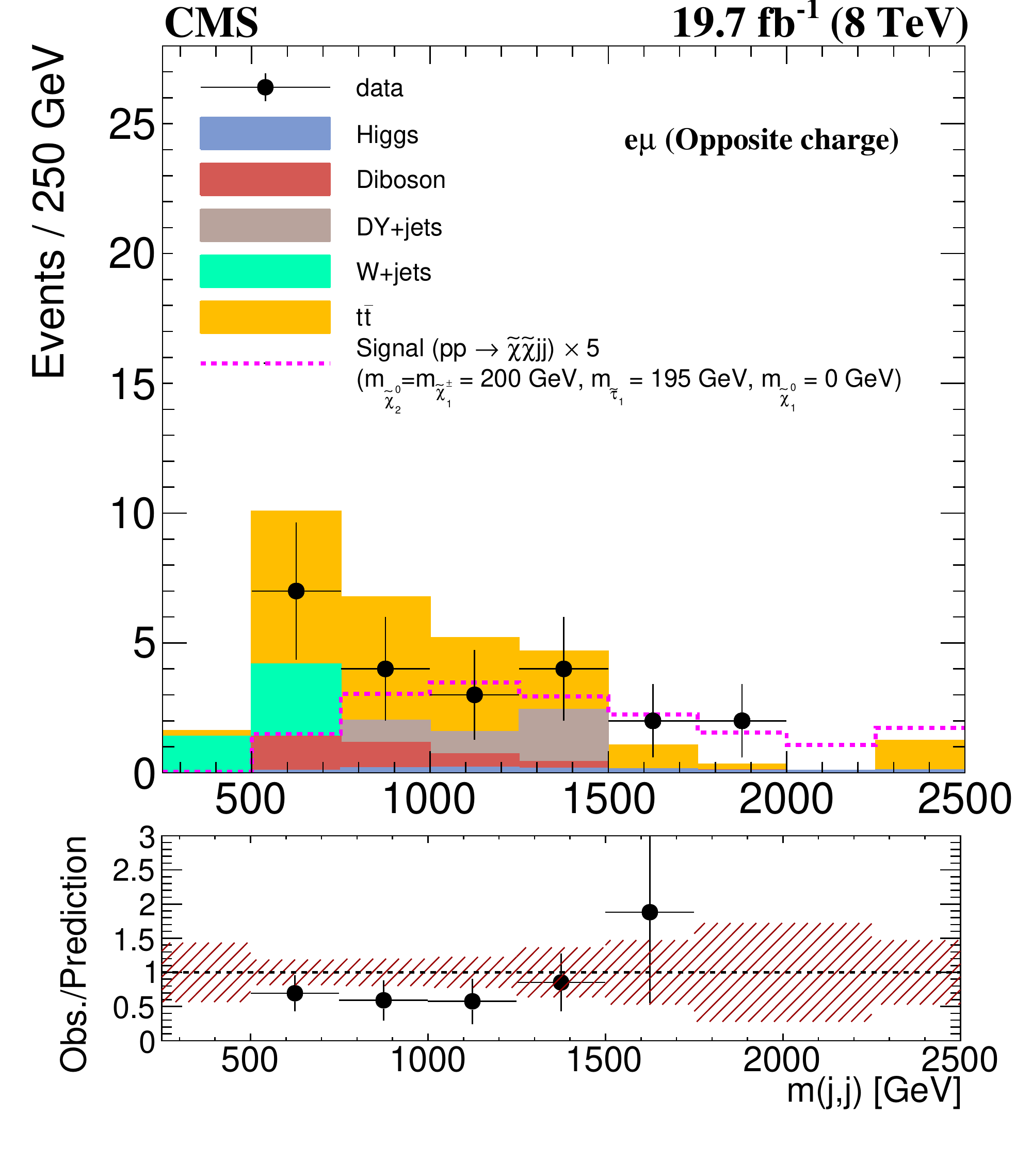}
    \includegraphics[width=0.45\textwidth]{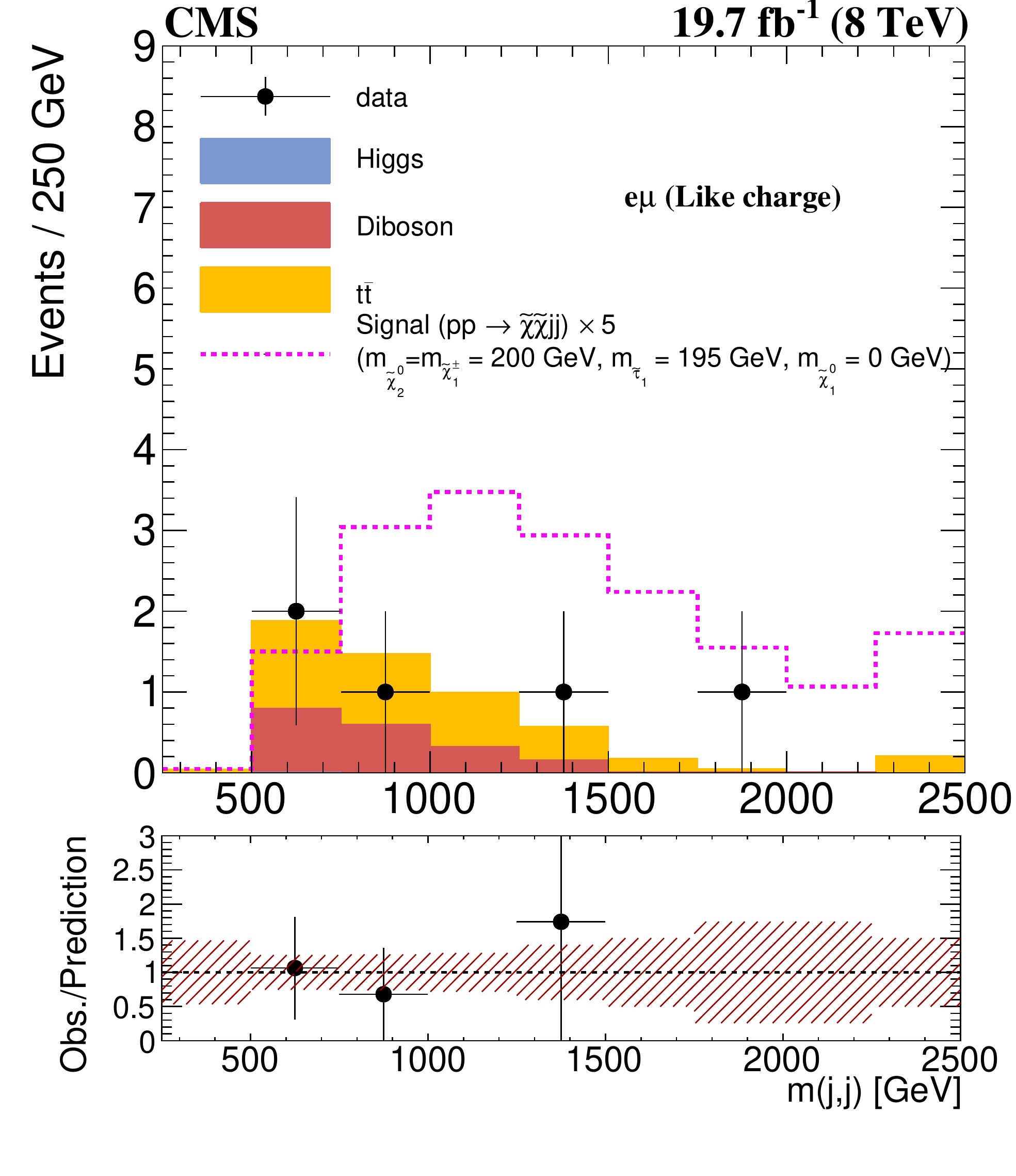}
  \caption{Dijet invariant mass distributions in the (upper left) OS $\mu\mu$, (upper right) SS $\mu\mu$, (lower left) OS $\Pe\mu$, and (lower right) SS e$\mu$ signal regions.
The signal scenario with $m_{\PSGczDt}=m_{\PSGcpmDo}=200$\GeV, $m_{\tilde{\tau}} = 195$\GeV, and $m_{\PSGczDo}=0$\GeV, as described in Section~\ref{sec:backgrounds}, is shown.
The signal events are scaled up by a factor of 5 for purposes of visibility. The shaded band in the ratio plot includes the systematic and statistical uncertainties in the background prediction.}
  \label{fig:SRPlots}
\end{figure}

\begin{figure}[tbh!]
  \centering
    \includegraphics[width=0.45\textwidth]{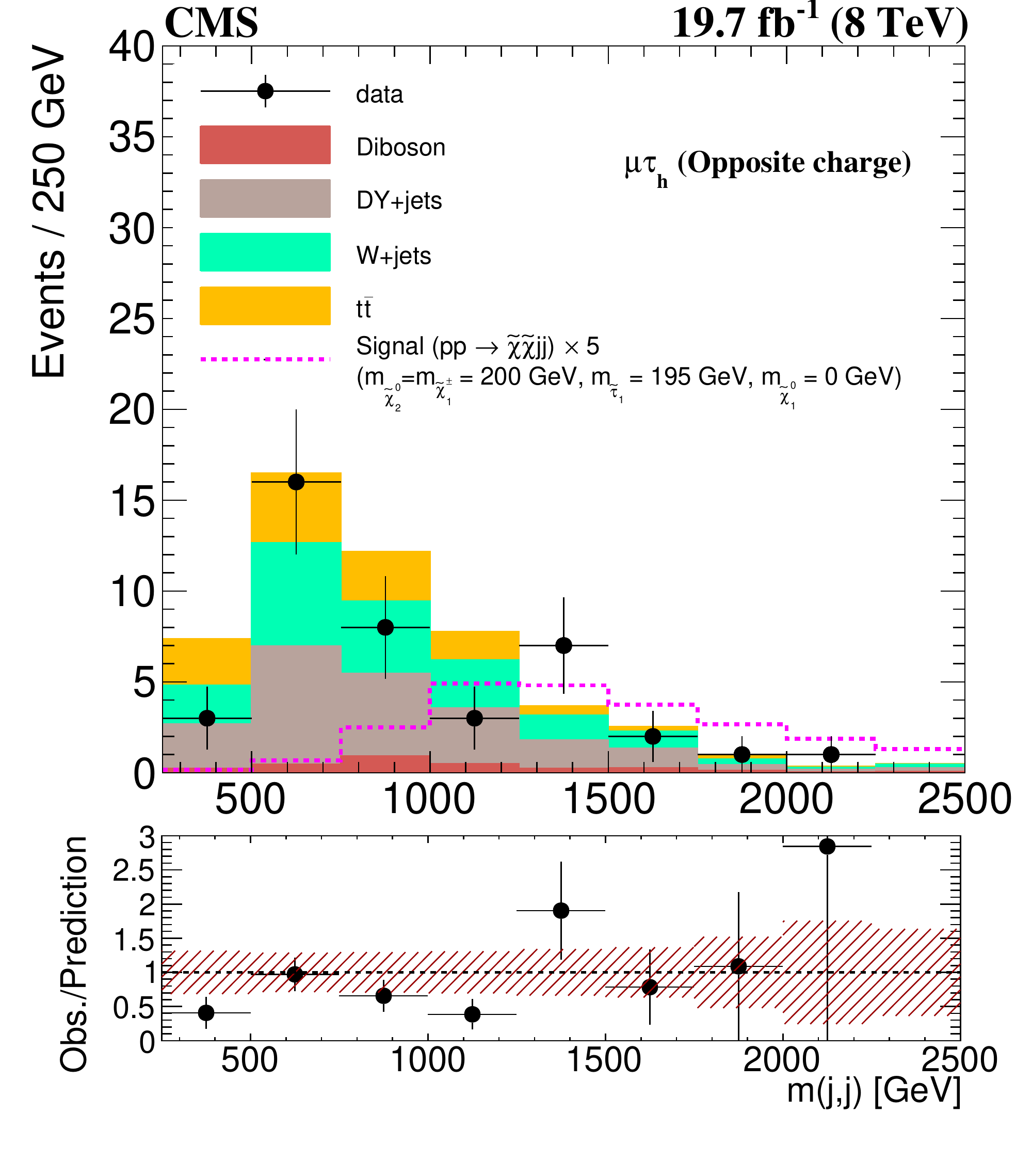}
    \includegraphics[width=0.45\textwidth]{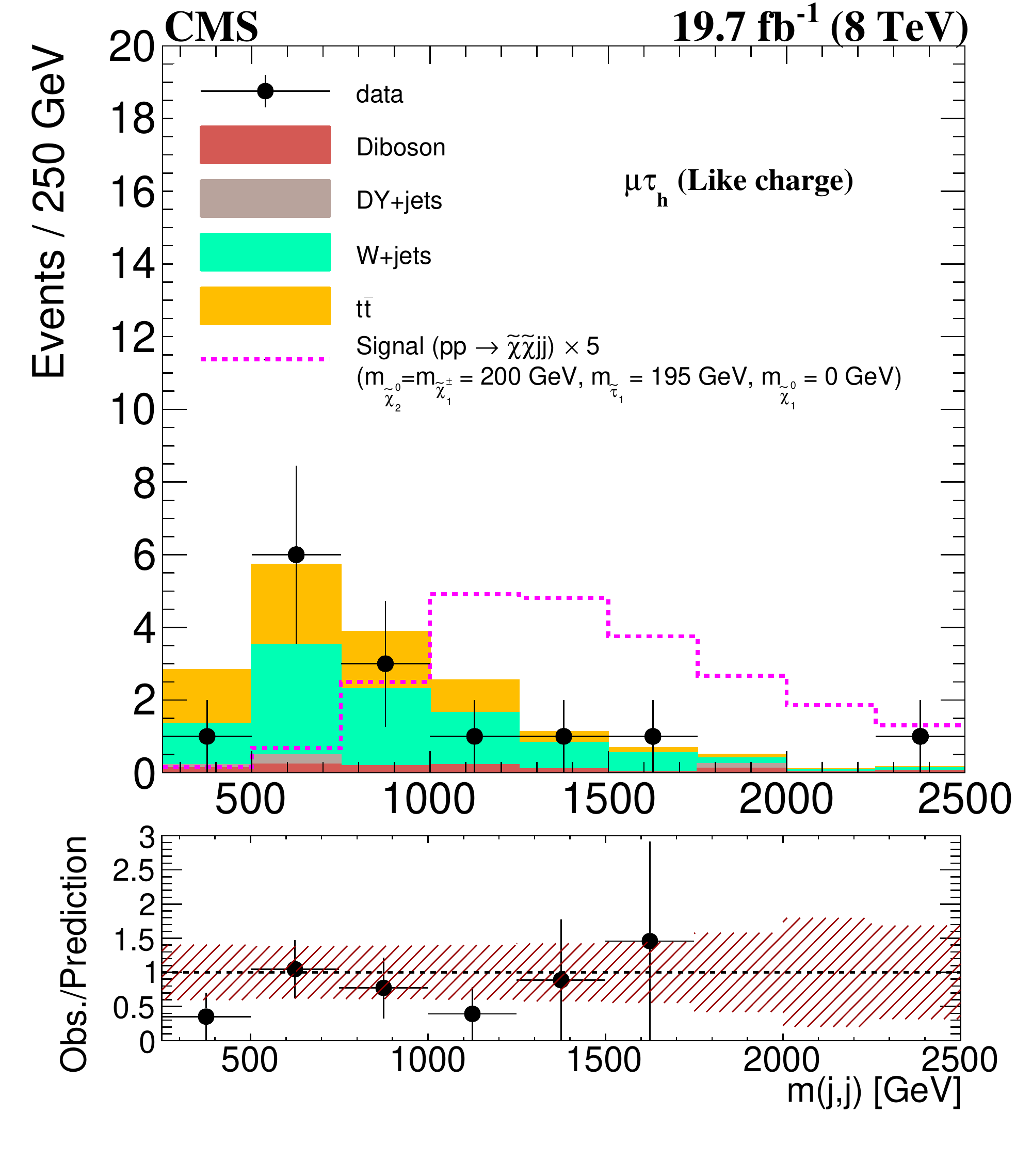}
    \includegraphics[width=0.45\textwidth]{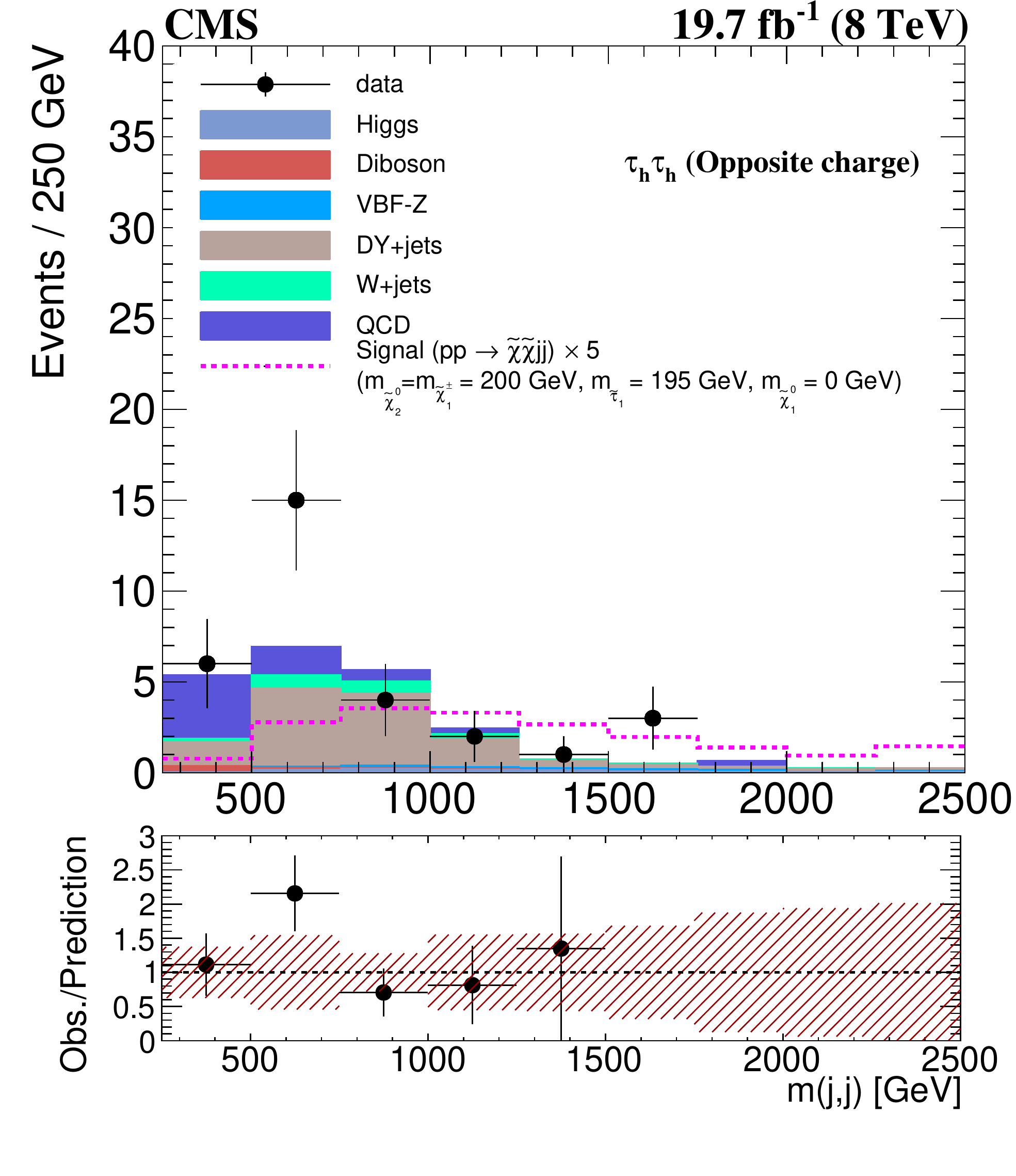}
    \includegraphics[width=0.45\textwidth]{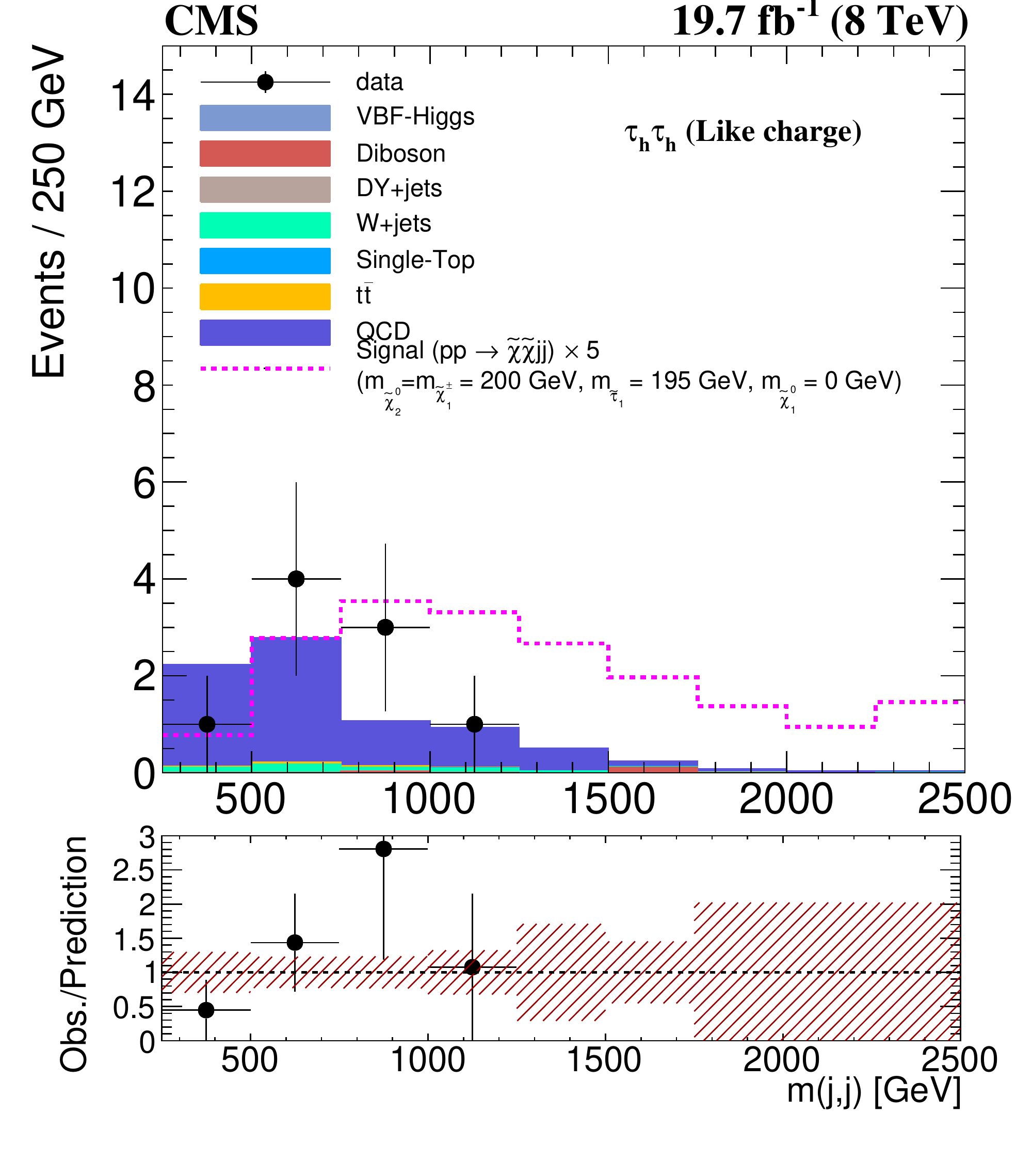}
  \caption{Dijet invariant mass distributions in the (upper left) OS $\mu\tau_{\tauh}$, (upper right) SS $\mu\tau_{\tauh}$, (lower left) OS
$\tau_{\tauh}\tau_{\tauh}$, and (lower right) SS $\tau_{\tauh}\tau_{\tauh}$ signal regions.
The signal scenario with $m_{\PSGczDt}=m_{\PSGcpmDo}=200$\GeV, $m_{\tilde{\tau}} = 195$\GeV, and $m_{\PSGczDo}=0$\GeV, as described in Section~\ref{sec:backgrounds}, is shown.
The signal events are scaled up by a factor of 5 for purposes of visibility. The shaded band in the ratio plot includes the systematic and statistical uncertainties in the background prediction.}
  \label{fig:SRPlots2}
\end{figure}

\begin{figure}[tbh!]
  \centering
    \includegraphics[width=0.45\textwidth]{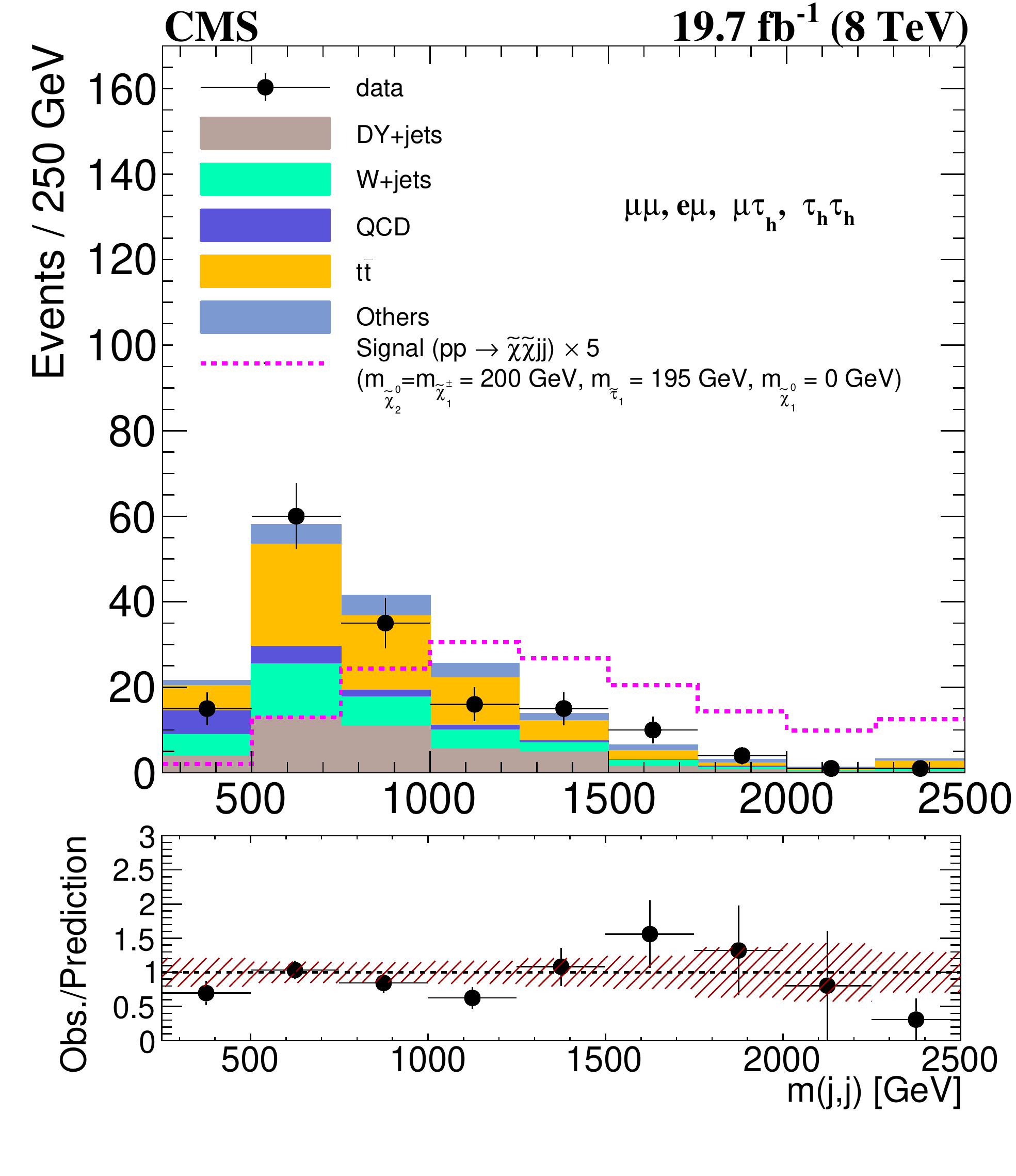}
  \caption{Dijet invariant mass distribution for the combination of all search channels.
The signal scenario with $m_{\PSGczDt}=m_{\PSGcpmDo}=200$\GeV, $m_{\tilde{\tau}} = 195$\GeV, and $m_{\PSGczDo}=0$\GeV, as described in Section~\ref{sec:backgrounds}, is shown.
The signal events are  scaled up by a factor of 5 for purposes of visibility. The shaded band in the ratio plot includes the systematic and statistical uncertainties in the background prediction.
}
  \label{fig:muTauCRs}
\end{figure}

Figures~\ref{fig:SRPlots} and~\ref{fig:SRPlots2} present the data in comparison to the predicted SM background. The combined results from all channels are shown in Fig.~\ref{fig:muTauCRs}.
Numerical results are given in Tables~\ref{table:expectations_OS} and \ref{table:expectations_SS}.
The observed numbers of events are seen to be consistent with the expected SM background in all search regions. Therefore the search does not reveal any evidence for new physics.

\begin{table*}[htbp]
  \topcaption{Number of observed events and corresponding background predictions for the OS channels.
The uncertainties are statistical, including the statistical uncertainties from the control regions and simulated event samples.}
  \centering{
  \begin{tabular}{ l | c  c  c  c }\hline
    Process  & $\mu^{\pm}\mu^{\mp}jj$ & $\Pe^{\pm}\mu^{\mp}jj$ & $\mu^{\pm}\tauh^{\mp}jj$ & $\tauh^{\pm}\tauh^{\mp}jj$ \\
[0.5ex] \hline
    \Z{}+jets       & $4.3 \pm 1.7$    & $3.7^{+2.1}_{-1.9}$ & $19.9 \pm 2.9$  & $12.3 \pm 4.4$\\
    \PW+jets        & $<$0.1         & $4.2 ^{+3.3}_{-2.5}$ & $17.3 \pm 3.0$  & $2.0  \pm 1.7$\\
    VV              & $2.8 \pm 0.5$  & $3.1  \pm 0.7$           & $2.9  \pm 0.5$  & $0.5  \pm 0.2$ \\
    $\ttbar$      & $24.0 \pm 1.7$ & $19.0 ^{+2.3}_{-2.4}$ & $11.7 \pm 2.8$    & -- \\
    QCD             & \NA               & \NA                     & \NA              & $6.3  \pm 1.8$ \\
    Higgs boson          & $1.0 \pm 0.1$  & $1.1  \pm 0.5$           & \NA              & $1.1  \pm 0.1$\\
    VBF \Z           & \NA               & \NA                     & \NA              & $0.7  \pm 0.2$ \\
    Total           & $32.2 \pm 2.4$ & $31.1 ^{+4.6}_{-4.1}$   & $51.8 \pm 5.1$  & $22.9 \pm 5.1$\\\hline
    Observed        &  31              & 22                     & 41              & 31\\
  \hline
  \end{tabular}
  }
  \label{table:expectations_OS}
\end{table*}

\begin{table*}[htbp]
  \topcaption{Number of observed events and corresponding background predictions for the SS channels.
The uncertainties are statistical, including the statistical uncertainties from the control regions and simulated event samples.}
  \centering{
  \begin{tabular}{ l | c  c  c  c }\hline
    Process  & $\mu^{\pm}\mu^{\pm}jj$ & $\Pe^{\pm}\mu^{\pm}jj$ & $\mu^{\pm}\tau_{\tauh}^{\pm}jj$ & $\tau_{\tauh}^{\pm}\tau_{\tauh}^{\pm}jj$ \\
[0.5ex] \hline
    \Z{}+jets       & $<$0.1 & $0 ^{+1.7}_{-0}$    & $0.5 \pm 0.2$  & $<$0.1\\
    \PW+jets        & $<$0.1     & $0 ^{+3.0}_{-0}$    & $9.3 \pm 2.3$  & $0.5 \pm 0.1$\\
    VV              & $2.1 \pm 0.3$          & $1.9 ^{+0.4}_{-0.2}$  & $1.1 \pm 0.2$  & $0.1 \pm 6.5 \times 10^{-2}$\\
    $\ttbar$      & $3.1 \pm 0.1$          & $3.5 ^{+0.7}_{-0.9}$  & $6.7 \pm 2.8$  & $0.1 \pm 1.2 \times 10^{-2}$\\
    Single top      & \NA        & \NA                     & \NA             & $<$0.1\\
    QCD             & \NA        & \NA                     & \NA             & $7.6 \pm 0.9$\\
    Higgs boson          & \NA        & \NA                     & \NA             & $<$0.1\\
    Total           & $5.4 \pm 0.3$          & $5.4 \pm^{3.5}_{0.9}$  & $17.6 \pm 3.8$ & $8.4 \pm 0.9 $\\\hline
    Observed        &  4         & 5                    &  14            & 9 \\
  \hline
  \end{tabular}
  }
  \label{table:expectations_SS}
\end{table*}

To quantify the sensitivity of this search, the results are interpreted in the context of the R-parity conserving minimal supersymmetric SM by considering production of charginos and neutralinos with two associated jets, as described in Section~\ref{sec:backgrounds}.
Models with a bino-like
$\PSGczDo$ and wino-like $\PSGczDt$ and $\PSGcpmDo$ are considered. Since the $\PSGczDt$ and $\PSGcpmDo$ belong to the same gauge
group multiplet, we set $m_{\PSGczDt}=m_{\PSGcpmDo}$ and present results as a function of this common mass and the LSP mass
$m_{\PSGczDo}$. In the presence of a light slepton,
$\tilde{\ell}=\tilde{\Pe}/\tilde{\mu}/\tilde{\tau}$, it is likely that the ${\PSGcpmDo}$ will decay to $\ell\nu\PSGczDo$
and the ${\PSGczDt}$ to $\ell^{+}\ell^{-}\PSGczDo$.
The results are interpreted by considering $\tilde{\ell}=\tilde{\tau}$ and assuming branching fractions
$\mathcal{B}(\PSGczDt\to \tau \tilde{\tau} \to \tau\tau \PSGczDo)=1$ and
$\mathcal{B}(\PSGcpmDo\to\nu\tilde{\tau} \to\nu\tau\PSGczDo)=1$.
To highlight how the VBF searches described in this paper complement other searches for
the electroweak production of SUSY particles~\cite{CMSEWK,Aad:2014nua}, two scenarios are considered: \textit{(i)} $m_{\PSGczDo}=0$\GeV (uncompressed-mass spectrum) and \textit{(ii)} $m_{\PSGcpmDo} - m_{\PSGczDo} = 50$\GeV (compressed-mass spectrum).

The cumulative signal event acceptance is shown in Table~\ref{ta:sigEff} at three stages of the analysis: accounting for the branching fractions for the SUSY event to yield the indicated two-lepton channel (BF), the acceptance following application of the central selection (Central), and the acceptance following the VBF selection (VBF).
The average $\pt$ values of the \Pe, $\mu$, and $\tauh$ objects in signal events are relatively soft, because of the energy and momentum carried by the associated neutrinos in the $\tau$ decays.
The OS and SS channels have similar signal acceptance because lepton pairs satisfying the event selection do not necessarily originate from the $\PSGczDt$ or $\PSGcpmDo \widetilde{\chi}^{\mp}_{1}$ decays.
The best signal sensitivity comes from the SS $\mu\mu$ and $\Pe \mu$ channels due to a better background suppression with respect to a given signal acceptance.

The expected signal yields from simulation with $m_{\PSGczDo} = 0$\GeV and
$\Delta m(\PSGcpmDo, \PSGczDo ) = 50$\GeV, are presented in Table~\ref{table:Signal_expectations}.
The signal acceptance depends on the mass $m_{\tilde{\tau}}$ of the intermediate $\tau$ slepton.
The results in Table~\ref{table:Signal_expectations} are presented under two different assumptions for $m_{\tilde{\tau}}$:
\textit{(i)} a fixed-mass difference assumption $\Delta m(\PSGcpmDo, \tilde{\tau}) = 5$\GeV, and
\textit{(ii)} an average-mass assumption  $m_{\tilde{\tau}} = 0.5 m_{\PSGcpmDo} + 0.5 m_{\PSGczDo}$.
In the compressed-mass-spectrum scenario, for which $m_{\PSGcpmDo} - m_{\PSGczDo} = 50$\GeV, the average-mass assumption yields significantly lower average lepton \pt than the fixed-mass assumption, and the acceptance is lower by a factor of 2--3.
In the uncompressed-mass-spectrum scenario, with $m_{\PSGczDo}$=0\GeV, the average-mass assumption produces larger average lepton \pt than the fixed-mass assumption, yielding an event acceptance that is 1.3--1.8 times larger.

\begin{table}[tbh!]
\centering
\topcaption
{Cumulative signal event acceptance after application of the BF, central, and VBF requirements (see text).
Note that the jet \pt threshold for the $\mu\mu jj$ and $\tauh\tauh jj$ final states is 30\GeV, while it is 50\GeV for the other final states.
}
\begin{tabular}{  c | c  c  c }
\hline

  Channel & BF($\ge$1$\ell_{1}$ \& $\ge$1$\ell_{2}$) &  Central & VBF \\
\hline
$\mu^{\pm}\tau_{\tauh}^{\mp}$ ($\mu^{\pm}\tau_{\tauh}^{\pm}$) & 0.399 & 0.020 (0.020) & 0.007 (0.007) \\
$\Pe^{\pm}\mu^{\mp}$ ($\Pe^{\pm}\mu^{\pm}$)                         & 0.152 & 0.037 (0.037) & 0.014 (0.014) \\
$\tau_{\tauh}^{\pm}\tau_{\tauh}^{\mp}$ ($\tau_{\tauh}^{\pm}\tau_{\tauh}^{\pm}$)     & 0.717 & 0.010 (0.010) & 0.009 (0.009) \\
$\mu^{\pm}\mu^{\mp}$ ($\mu^{\pm}\mu^{\pm}$)                         & 0.081 & 0.018 (0.018) & 0.007 (0.017) \\
\hline \end{tabular}
\label{ta:sigEff}
\end{table}

\begin{table*}[htbp]
  \topcaption{Signal event yields from simulation. The first terms $\{m_{\PSGcpmDo}, m_{\tilde{\tau}} \}$ correspond to the fixed-mass difference assumption $\Delta m(\PSGcpmDo, \tilde{\tau}) = 5$\GeV, while the terms in parentheses
($\{m_{\PSGcpmDo}, m_{\tilde{\tau}}\}$) correspond to the average-mass assumption $m_{\tilde{\tau}} = 0.5m_{\PSGczDo} + 0.5 m_{\PSGcpmDo}$.}
  \centering
  \begin{tabular}{ l | c  c  c  c  c }\hline
    \{$m(\PSGcpmDo)$,$m(\tilde{\tau})$\} [\GeVns{}] & $\mu^{\pm}\mu^{\pm}jj$ (loose) & $\mu^{\pm}\mu^{\mp}jj $ (tight)  & $\Pe\mu jj$ & $\mu\tauh jj$ &
$\tauh\tauh jj$ \\[0.5ex] \hline
    \multicolumn{6}{c}{$m(\PSGczDo) = 0$\GeV}\\[0.5ex] \hline
    \{100, 95\} (\{100, 50\}) & $16 (29)$ & $6.6 (12)$ & $13 (24)$ & $7.1 (9.4)$ & $8.7 (10.7)$ \\
    \{200, 195\} (\{200, 100\}) & $5.4 (9.7)$ & $1.8 (3.1)$ & $3.5 (6.3)$ & $4.5 (6.0)$ & $3.8 (4.7)$ \\
    \{300, 295\} (\{300, 150\}) & $2.3 (4.1)$ & $0.68 (1.2)$ & $1.4 (2.4)$ & $1.9 (2.5)$ & $1.5 (2.0)$\\
    \{400, 395\} (\{400, 200\}) & $0.57 (1.0)$ & $0.17 (0.30)$ & $0.35 (0.62)$ & $0.46 (0.63)$ & $0.38 (0.51)$\\
  \hline
  \multicolumn{6}{c}{$\Delta m(\PSGcpmDo - \PSGczDo ) = 50$\GeV}\\ [0.5ex] \hline
    \{200, 195\} (\{200, 175\}) & $1.4 (0.5)$ & $0.85 (0.33)$ & $1.7 (0.65)$ & $0.99 (0.35)$ & $0.46 (0.09)$\\
    \{300, 295\} (\{300, 275\}) & $0.47 (0.18)$ & $0.28 (0.11)$ & $0.58 (0.23)$ & $0.40 (0.14)$ & $0.20 (0.04)$\\
    \{400, 395\} (\{400, 375\}) & $0.12 (0.05)$ & $0.08 (0.03)$ & $0.15 (0.06)$ & $0.10 (0.03)$ & $0.05 (0.01)$\\
  \hline
  \end{tabular}
\label{table:Signal_expectations}
\end{table*}

The calculation of the exclusion limit is obtained by using the $m_{jj}$ distribution in each channel to construct a combined
likelihood in bins of $m_{jj}$ and computing a 95\% confidence level (CL)
upper limit on the signal cross section using the asymptotic CL$_{s}$ criterion \cite{CLs1, Junk, Cowan:2010js}.
Systematic uncertainties are taken into account as nuisance parameters, which are removed by marginalization, assuming a gamma or
log-normal prior for normalization parameters, and Gaussian priors for mass spectrum shape uncertainties.
The combination of the eight search channels requires simultaneous analysis of the
data from the individual channels, accounting for all statistical and systematic uncertainties
and their correlations.
Correlations among backgrounds, both within a channel and across channels, are taken into consideration in the limit calculation.
For example, the uncertainties in physics object identification and reconstruction are treated as correlated for channels with a common particle in their final states, while the uncertainty in the integrated luminosity is treated as correlated across channels.
The uncertainties resulting from the number of simulated events, and from the event acceptance variation with different sets of PDFs in a given $m_{jj}$ bin, are treated as uncorrelated within a channel and correlated across channels.
The uncertainties due to the closure tests are treated as uncorrelated within and across the different final states.

\begin{figure}[tbh]
   \centering
    \includegraphics[width=.45\textwidth]{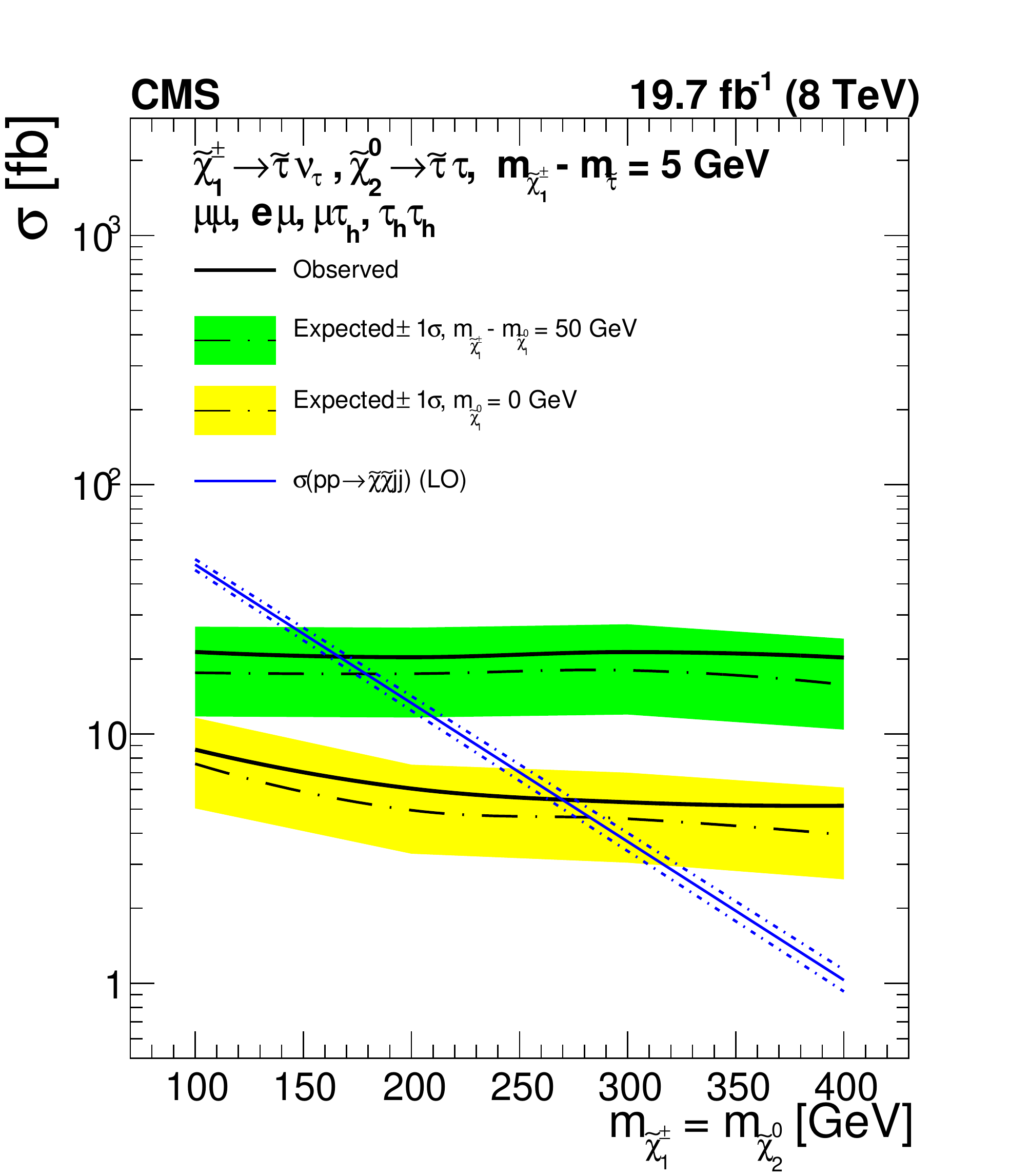}
    \includegraphics[width=.45\textwidth]{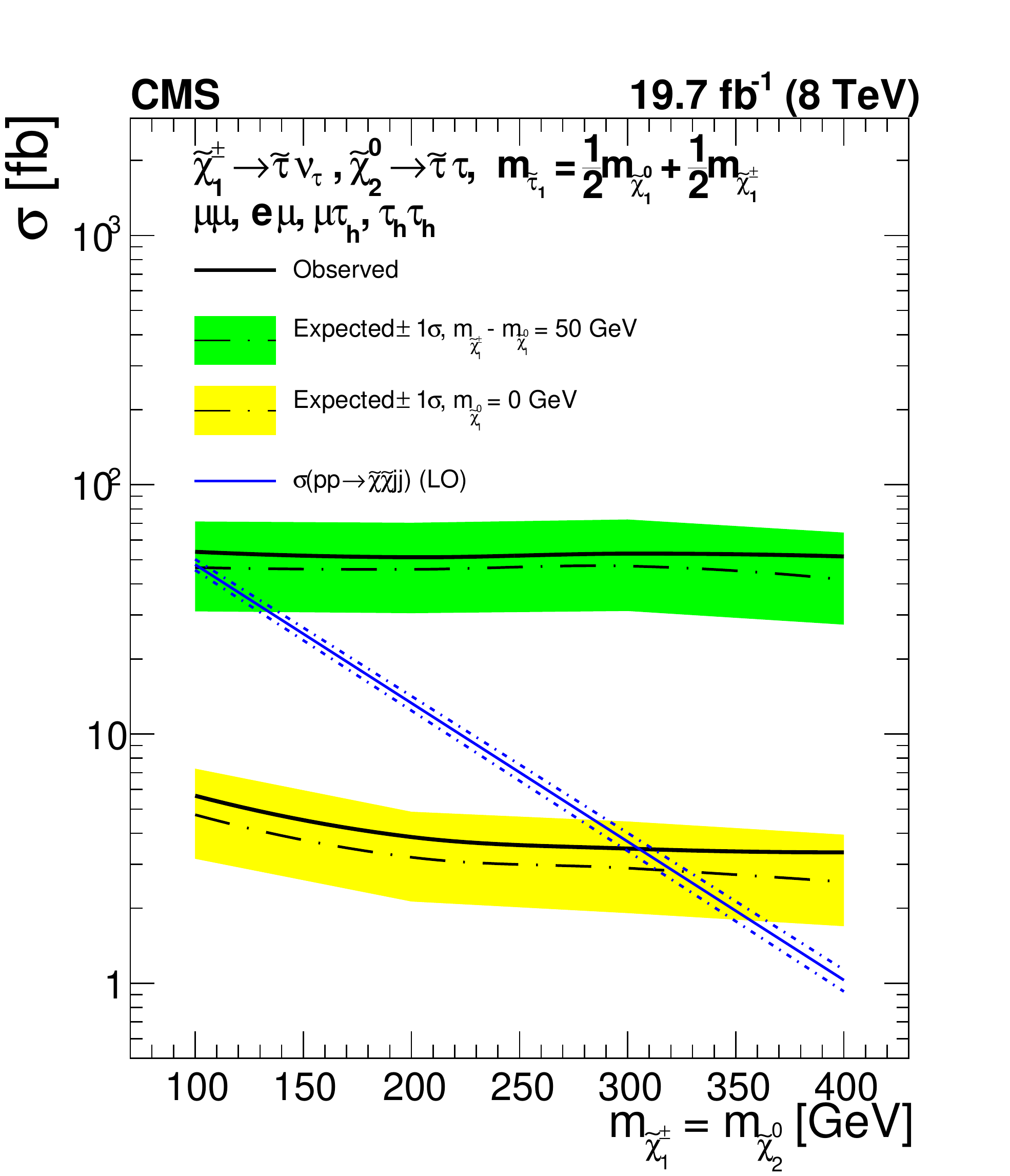}
     \caption{Combined 95\% CL upper limits on the cross section as a function of $m_{\PSGczDt}=m_{\PSGcpmDo}$. The signal cross section is calculated with the VBF jet selection: jet \pt $>$ 30\GeV, $\abs{\Delta\eta(\text{jets})} > 4.2$, and $\eta_{\text{1}} \eta_{\text{2}} < 0$.
(left) The results for the fixed-mass difference assumption, in which $m_{\PSGcpmDo} - m_{\tilde{\tau}}= 5$\GeV, for $m_{\PSGcpmDo} - m_{\PSGczDo} = 50$\GeV (compressed-mass spectrum) and $m_{\PSGczDo} = 0$\GeV (uncompressed-mass spectrum). (right) The corresponding results for the average-mass assumption, in which $m_{\tilde{\tau}} = 0.5 m_{\PSGcpmDo} + 0.5 m_{\PSGczDo}$.}
 \label{fig:combinedLimit}
\end{figure}

Figures~\ref{fig:combinedLimit} (left) and~\ref{fig:combinedLimit} (right) show the expected and observed limits as well as the theoretical cross section as functions of $m_{\PSGcpmDo}$ for, respectively, the fixed- and average-mass $m_{\tilde{\tau}}$ assumptions.
For the fixed-mass assumption with a compressed-mass spectrum ($m_{\PSGcpmDo} -m_{\PSGczDo} = 50\GeV$), $\PSGczDt$ and $\PSGcpmDo$ gauginos with masses below 170\GeV are excluded, where the previous ATLAS and CMS SUSY searches do not probe.
For the average-mass assumption with an uncompressed-mass spectrum ($m_{\PSGczDo} = 0$), the corresponding limit is 300\GeV.
These mass limits are conservatively determined using the theoretical cross section minus its one standard deviation uncertainty.
The $m_{\PSGcpmDo}$ limits of 320 and 380\GeV for $m_{\PSGczDo}$ = 0\GeV in Refs.~\cite{CMSEWK, Aad:2014nua} can be compared to the corresponding result of 300\GeV in the present analysis [see the yellow band in Fig.~\ref{fig:combinedLimit} (right)].

\section{Summary}
A search is presented for non-coloured supersymmetric particles in the vector-boson fusion (VBF) topology
using data corresponding to an integrated luminosity of 19.7\fbinv collected with the CMS detector
in proton-proton collisions at $\sqrt{s}$ = 8\TeV.
This is the first search for SUSY in the VBF topology.
The search utilizes events in eight different final states covering both same- and opposite-sign dilepton pairs.
The leptons considered are electrons, muons, and hadronically decaying $\tau$ leptons.
The VBF topology requires two well-separated jets that appear in opposite hemispheres, with large invariant mass $m_{jj}$.
The observed $m_{jj}$ distributions do not reveal any evidence for new physics.
The results are used to exclude a range of $\PSGcpmDo$ and $\PSGczDt$ gaugino masses.
For models in which the $\PSGczDo$ lightest-supersymmetric-particle mass is zero, and in which the $\PSGcpmDo$ and $\PSGczDt$ branching fractions to $\tau$ leptons are large, $\PSGcpmDo$ and $\PSGczDt$ masses up to 300\GeV are excluded at 95\% CL.
For a compressed-mass-spectrum scenario, in which $m_{\PSGcpmDo} -m_{\PSGczDo} = 50$\GeV, the corresponding limit is 170\GeV.
While many previous studies at the LHC have focused on strongly coupled supersymmetric particles, including searches for charginos and neutralinos produced in gluino or squark decay chains, and a number of studies have presented limits on the Drell-Yan production of charginos and neutralinos, this analysis obtains the most stringent limits to date on the production of charginos and neutralinos decaying to $\tau$ leptons
in compressed-mass-spectrum scenarios defined by the mass separation $\Delta m = m_{\PSGcpmDo} - m_{\PSGczDo} < 50$\GeV.

\begin{acknowledgments}
\hyphenation{Bundes-ministerium Forschungs-gemeinschaft Forschungs-zentren} We congratulate our colleagues in the CERN accelerator departments for the excellent performance of the LHC and thank the technical and administrative staffs at CERN and at other CMS institutes for their contributions to the success of the CMS effort. In addition, we gratefully acknowledge the computing centres and personnel of the Worldwide LHC Computing Grid for delivering so effectively the computing infrastructure essential to our analyses. Finally, we acknowledge the enduring support for the construction and operation of the LHC and the CMS detector provided by the following funding agencies: the Austrian Federal Ministry of Science, Research and Economy and the Austrian Science Fund; the Belgian Fonds de la Recherche Scientifique, and Fonds voor Wetenschappelijk Onderzoek; the Brazilian Funding Agencies (CNPq, CAPES, FAPERJ, and FAPESP); the Bulgarian Ministry of Education and Science; CERN; the Chinese Academy of Sciences, Ministry of Science and Technology, and National Natural Science Foundation of China; the Colombian Funding Agency (COLCIENCIAS); the Croatian Ministry of Science, Education and Sport, and the Croatian Science Foundation; the Research Promotion Foundation, Cyprus; the Ministry of Education and Research, Estonian Research Council via IUT23-4 and IUT23-6 and European Regional Development Fund, Estonia; the Academy of Finland, Finnish Ministry of Education and Culture, and Helsinki Institute of Physics; the Institut National de Physique Nucl\'eaire et de Physique des Particules~/~CNRS, and Commissariat \`a l'\'Energie Atomique et aux \'Energies Alternatives~/~CEA, France; the Bundesministerium f\"ur Bildung und Forschung, Deutsche Forschungsgemeinschaft, and Helmholtz-Gemeinschaft Deutscher Forschungszentren, Germany; the General Secretariat for Research and Technology, Greece; the National Scientific Research Foundation, and National Innovation Office, Hungary; the Department of Atomic Energy and the Department of Science and Technology, India; the Institute for Studies in Theoretical Physics and Mathematics, Iran; the Science Foundation, Ireland; the Istituto Nazionale di Fisica Nucleare, Italy; the Ministry of Science, ICT and Future Planning, and National Research Foundation (NRF), Republic of Korea; the Lithuanian Academy of Sciences; the Ministry of Education, and University of Malaya (Malaysia); the Mexican Funding Agencies (CINVESTAV, CONACYT, SEP, and UASLP-FAI); the Ministry of Business, Innovation and Employment, New Zealand; the Pakistan Atomic Energy Commission; the Ministry of Science and Higher Education and the National Science Centre, Poland; the Funda\c{c}\~ao para a Ci\^encia e a Tecnologia, Portugal; JINR, Dubna; the Ministry of Education and Science of the Russian Federation, the Federal Agency of Atomic Energy of the Russian Federation, Russian Academy of Sciences, and the Russian Foundation for Basic Research; the Ministry of Education, Science and Technological Development of Serbia; the Secretar\'{\i}a de Estado de Investigaci\'on, Desarrollo e Innovaci\'on and Programa Consolider-Ingenio 2010, Spain; the Swiss Funding Agencies (ETH Board, ETH Zurich, PSI, SNF, UniZH, Canton Zurich, and SER); the Ministry of Science and Technology, Taipei; the Thailand Center of Excellence in Physics, the Institute for the Promotion of Teaching Science and Technology of Thailand, Special Task Force for Activating Research and the National Science and Technology Development Agency of Thailand; the Scientific and Technical Research Council of Turkey, and Turkish Atomic Energy Authority; the National Academy of Sciences of Ukraine, and State Fund for Fundamental Researches, Ukraine; the Science and Technology Facilities Council, UK; the US Department of Energy, and the US National Science Foundation.

Individuals have received support from the Marie-Curie programme and the European Research Council and EPLANET (European Union); the Leventis Foundation; the A. P. Sloan Foundation; the Alexander von Humboldt Foundation; the Belgian Federal Science Policy Office; the Fonds pour la Formation \`a la Recherche dans l'Industrie et dans l'Agriculture (FRIA-Belgium); the Agentschap voor Innovatie door Wetenschap en Technologie (IWT-Belgium); the Ministry of Education, Youth and Sports (MEYS) of the Czech Republic; the Council of Science and Industrial Research, India; the HOMING PLUS programme of the Foundation for Polish Science, cofinanced from European Union, Regional Development Fund; the OPUS programme of the National Science Center (Poland); the Compagnia di San Paolo (Torino); the Consorzio per la Fisica (Trieste); MIUR project 20108T4XTM (Italy); the Thalis and Aristeia programmes cofinanced by EU-ESF and the Greek NSRF; the National Priorities Research Program by Qatar National Research Fund; the Rachadapisek Sompot Fund for Postdoctoral Fellowship, Chulalongkorn University (Thailand); and the Welch Foundation, contract C-1845.
\end{acknowledgments}

\bibliography{auto_generated}

\cleardoublepage \appendix\section{The CMS Collaboration \label{app:collab}}\begin{sloppypar}\hyphenpenalty=5000\widowpenalty=500\clubpenalty=5000\textbf{Yerevan Physics Institute,  Yerevan,  Armenia}\\*[0pt]
V.~Khachatryan, A.M.~Sirunyan, A.~Tumasyan
\vskip\cmsinstskip
\textbf{Institut f\"{u}r Hochenergiephysik der OeAW,  Wien,  Austria}\\*[0pt]
W.~Adam, E.~Asilar, T.~Bergauer, J.~Brandstetter, E.~Brondolin, M.~Dragicevic, J.~Er\"{o}, M.~Flechl, M.~Friedl, R.~Fr\"{u}hwirth\cmsAuthorMark{1}, V.M.~Ghete, C.~Hartl, N.~H\"{o}rmann, J.~Hrubec, M.~Jeitler\cmsAuthorMark{1}, V.~Kn\"{u}nz, A.~K\"{o}nig, M.~Krammer\cmsAuthorMark{1}, I.~Kr\"{a}tschmer, D.~Liko, T.~Matsushita, I.~Mikulec, D.~Rabady\cmsAuthorMark{2}, B.~Rahbaran, H.~Rohringer, J.~Schieck\cmsAuthorMark{1}, R.~Sch\"{o}fbeck, J.~Strauss, W.~Treberer-Treberspurg, W.~Waltenberger, C.-E.~Wulz\cmsAuthorMark{1}
% ...
\vskip\cmsinstskip
% ...
1:~~Also at Vienna University of Technology, Vienna, Austria\\
2:~~Also at CERN, European Organization for Nuclear Research, Geneva, Switzerland\\
% ...

\end{sloppypar}
\end{document}